\documentstyle[12pt]{article}

\newcommand{\be}{\begin{equation}}
\newcommand{\ee}{\end{equation}}
\newcommand{\ba}{\begin{eqnarray}}
\newcommand{\ea}{\end{eqnarray}}
\newcommand{\n}{\nonumber \\}

\newcommand{\eq}[1]{(\ref{#1})}
\newcommand{\sfrac}[2]{{\textstyle \frac{#1}{#2}}}
\newcommand{\ket}[1]{|#1\rangle}
\newcommand{\bra}[1]{\langle#1|}

\newcommand\bZ{\bf Z}

\newcommand\no{\noindent}

\newcommand\x[2]{x^{#1}_{#2}}

\newcommand\laa[1]{\lambda^{(#1)}}
\newcommand\Sca[3]{{1\over#3!}\oint\prod_{#2=1}^{#3}{d#1_#2\over 2\pi i#1_#2}}

\newcommand\res[3]{		   \prod_{#2=1}^{#3}{d#1_#2\over 2\pi i#1_#2}}

\newcommand\sRangle[2]{\rangle'_{#1;#2}}
\newcommand\Daa[1]{\Delta(x^{#1})}			
\newcommand\Dab[2]{\Gamma(x^{#1},x^{#2})}		
\newcommand\tDab[2]{\widetilde\Gamma(x^{#1},x^{#2})}

\def\boxline#1{\vbox{\hrule\hbox{\vrule\vbox{#1}\vrule}\hrule}}
\def\boxNW#1{\vbox{\hrule\hbox{\vrule\vbox{#1}}}}
\def\boxES#1{\vbox{\hbox{\vbox{#1}\vrule}\hrule}}
\def\Between#1#2#3#4{ 
\raise-#1mm\vbox to#1mm{\hsize #2mm \vbox{\vskip #3mm\centerline{#4} } }
}
\def\Square#1#2#3#4{ 
\raise-#1mm\boxline{
\vbox to#1mm{\hsize #1mm \vbox{\vskip #2mm\no\hskip4pt {#3}
			       \vskip-#2mm\vskip-8pt\centerline{#4} }} }
}
\def\Young#1#2#3#4#5#6#7#8#9{
\raise-#9mm\boxNW{\vbox to#1mm{\hsize#6mm
	\vbox{\vskip#8mm\no\hskip#8mm$\;\lambda$} }}
\kern-#6mm
\raise-#1mm\boxES{\vbox to #5mm{\hsize #7mm $ $}}\kern-.4pt
\raise-#2mm\boxES{\vbox to #5mm{\hsize #7mm $ $}}\kern-.4pt
\raise-#3mm\boxES{\vbox to #5mm{\hsize #7mm $ $}}\kern-.4pt
\raise-#4mm\boxES{\vbox to #5mm{\hsize #7mm $ $}}\kern-.4pt
\raise-#5mm\boxES{\vbox to #5mm{\hsize #7mm $ $}}
}
\def\Galilei{
\Between{10}{15}3{${G}_s\; :\;$}
  \Young{10}8642{15}32{9.9}
\Between{10}{20}3{$\longmapsto$}
 \Square{14}7{$r$}{$s$} \kern-.4pt  
  \Young{10}8642{15}32{9.9}
\Between{10}{10}{10}{\quad .}
}
\def\Dummy{{{ }\over{ }}}
\def\generalYoung{
\Between{10}{15}7{$\lambda=$}
 \Square{20}9{$r_{n-1}$}{$s_{n-1}\Dummy$} \kern-.4pt
 \Square{17}8{$r_{n-2}$}{$s_{n-2}\Dummy$} \kern-.4pt
\Between{10}{15}7{$\cdots\cdots$}
 \Square{13}6{$r_2    $}{$\;s_2  \Dummy$} \kern-.4pt
 \Square{10}5{$r_1    $}{$\;s_1  \Dummy$} \kern-.4pt
\Between{15}{10}{10}{\hfill .}
}
\def\generalYoung{
\vskip.25cm
\noindent
\makebox[  4cm]{ }
\makebox[  2cm]{$s^1$}\hskip-.4pt
\makebox[1.7cm]{$s^2$}
\makebox[1.4cm]{ }
\makebox[1.4cm]{$s^{N-2}$}\hskip-.35pt
\makebox[1.3cm]{$s^{N-1}$}
\hfill\break
 \makebox[  4cm][r]{$\hfill\lambda=$}
\framebox[  2cm][l]{\rule[  -1cm]{0cm}{  2cm}$r^1$}\hskip-.4pt
\framebox[1.7cm][l]{\rule[-0.7cm]{0cm}{1.7cm}$r^2$}
 \makebox[1.4cm]			    {\raisebox{.25cm}{$\cdots\cdots$}}
\framebox[1.4cm][l]{\rule[-0.4cm]{0cm}{1.4cm}\raisebox{.25cm}{$r^{N-2}$}
							     }\hskip-.4pt
\framebox[1.3cm][l]{\rule[-0.2cm]{0cm}{1.2cm}\raisebox{.25cm}{$r^{N-1}$}}
\makebox[1cm][r]{.}
\vskip.3cm
}

\begin{document}

\renewcommand{\thefootnote}{\fnsymbol{footnote}}
\font\csc=cmcsc10 scaled\magstep1
{\baselineskip=14pt
 \rightline{
 \vbox{\hbox{RIMS-1009}
       \hbox{YITP/U-95-3}
       \hbox{SULDP-1995-2}
       \hbox{February 1995}
       \hbox{revised May 1995}
}}}

\vskip 5mm
\begin{center}
{\large\bf
Excited States of Calogero-Sutherland Model\\
\vskip 3mm
and Singular Vectors of the $W_N$ Algebra
}

\vspace{10mm}

{\csc Hidetoshi AWATA}\footnote{JSPS fellow}\setcounter{footnote}
{0}\renewcommand{\thefootnote}{\arabic{footnote}}\footnote{
      e-mail address : awata@kurims.kyoto-u.ac.jp},
{\csc Yutaka MATSUO}\footnote{
      e-mail address : yutaka@yukawa.kyoto-u.ac.jp},
{\csc Satoru ODAKE}\footnote{
      e-mail address : odake@yukawa.kyoto-u.ac.jp}\\
\vskip.1in
and \
{\csc Jun'ichi SHIRAISHI}$^*$\footnote{
      e-mail address : shiraish@danjuro.phys.s.u-tokyo.ac.jp}

{\baselineskip=15pt
\it\vskip.25in
  $^1$Research Institute for Mathematical Sciences \\
  Kyoto University, Kyoto 606, Japan \\
\vskip.1in
  $^2$Uji Research Center, Yukawa Institute for Theoretical Physics \\
  Kyoto University, Uji 611, Japan \\
\vskip.1in
  $^3$Department of Physics, Faculty of Liberal Arts \\
  Shinshu University, Matsumoto 390, Japan \\
\vskip.1in
  $^4$Department of Physics, Faculty of Science \\
  University of Tokyo, Tokyo 113, Japan
}

\end{center}

\vspace{4mm}

\begin{abstract}
{
Using the collective field method,
we find a relation between the Jack symmetric polynomials,
which describe the excited states of the Calogero-Sutherland model,
and the singular vectors of the $W_N$ algebra.
Based on this relation, we obtain their integral representations.
We also give a direct algebraic method which leads to the same result,
and integral representations of the skew-Jack polynomials.
}
\end{abstract}

hep-th/9503043
\setcounter{footnote}{0}
\renewcommand{\thefootnote}{\arabic{footnote}}
\newpage
\section{Introduction}

Calogero-Sutherland model \cite{rCS}
describes a system of non-relativistic
particles on a circle under the inverse square potential.
Its Hamiltonian and momentum are
\be
  H_{CS}=\sum_{j=1}^{N_0}\frac{1}{2}
  \biggl(\frac{1}{i}\frac{\partial}{\partial q_j}\biggr)^2
  +\Bigl(\frac{\pi}{L}\Bigr)^2
  \sum_{i,j=1 \atop i<j}^{N_0}
  \frac{\beta(\beta-1)}{\sin^2\frac{\pi}{L}(q_i-q_j)},
  \quad
  P_{CS}=\sum_{j=1}^{N_0}
  \frac{1}{i}\frac{\partial}{\partial q_j},
  \label{CS}
\ee
where $\beta$ is a coupling constant.
This model was introduced by Sutherland several years ago
and has been known to describe a system with the generalized
exclusion principle in $1+1$ dimension \cite{rHal}\cite{rHa}.
Recently, this model and its various cousins
(Haldane-Shastry models \cite{rHS} and similar models with internal
degree of freedom \cite{rKHH})
have been intensively studied.
Among many beautiful results, we may mention
Yangian symmetry \cite{rHHTBP}\cite{rBGHP},
$W_{1+\infty}$ symmetry \cite{rUWH},
and their relations with 2D Yang-Mills theory \cite{rGN}\cite{rMP1}
and the matrix models \cite{rAJ}\cite{rSLA}.

Among others, the development which is particularly
relevant to our study
may be the evaluation 
of the dynamical correlation functions \cite{rSLA}--\cite{rMP2}.
In these calculations, they essentially used
the mathematical properties of the Jack symmetric polynomial,
namely the eigenstates of the Calogero-Sutherland model,
developed by Stanley and Macdonald \cite{rS}.
To go further to get higher correlation functions,
it is desirable to obtain the explicit
expression of the Jack polynomial.
In this paper, we derive such formula as the
multiple-integrals which typically appeared in the conformal
field theory in the Coulomb-gas representation.

The Jack symmetric polynomial is a deformation of the Schur symmetric
polynomial ($\beta=1$ case)
which can be expressed in terms of a free fermion \cite{rDJKM}.
Natural questions arise ; does the Calogero-Sutherland
system have some field theoretical reformulation
in terms of free bosons?
In \cite{rAMOS}, we studied this problem and
obtained the collective field description \cite{rJS}
of the Calogero-Sutherland system.
In particular, the Hamiltonian becomes
cubic in free bosons and takes the following form,
\be
  \hat{H}_{\beta}  =
  \sqrt{2\beta}\sum_{n>0}a_{-n}L_n
  +\sum_{n>0}a_{-n}a_n(N_0\beta+\beta-1-\sqrt{2\beta}a_0).
\label{eHb}
\ee
Here $L_n$ is the Coulomb-gas representation
of the Virasoro generator whose central charge
is given by $1-\frac{6(1-\beta)^2}{\beta}$.
By using this Hamiltonian, we derived the
explicit form of some of the Jack symmetric polynomials.
We observed that the eigenstates for a single pseudo-particle
(hole) excitation
have an interpretation as the screening charges of
the Virasoro algebra.
We obtained also the integral representation of the Jack symmetric
polynomial with the rectangular Young diagram.
These observations shows that there are some relations
between Calogero-Sutherland model and the representation theory
of the Virasoro algebra.

In fact,  Mimachi and Yamada showed that the Virasoro
singular vectors are expressed in terms of the Jack symmetric
polynomial with the rectangular Young diagram \cite{rMY1}.
Their derivation is based on the direct
computation.  However, we may give
more instructive proof by our bosonized Hamiltonian \eq{eHb}.
Indeed, when the Hamiltonian is acted on the singular
vectors, the cubic term vanishes because of the highest
weight condition and the action of the bilinear term
is trivially diagonal. By this observation,
the Virasoro singular vector naturally becomes the
eigenstate of the Hamiltonian.

In this paper we generalize this result.
{\em When the Young diagram consists of $N-1$ rectangles,
the Jack symmetric polynomials with such Young diagram are related
to the singular vectors of the $W_N$ algebra.}

After projecting out some of the redundant degrees of freedoms,
we obtain the integral representations of
the Jack polynomials with arbitrary Young diagram \cite{rMY2}.

The nature of this formula is revealed by
decomposing the integral
into several alternating actions of two types of operators,
(a) $G_s$: the Galilean boost which amounts to
adding ($r,s$) rectangle to the original Young diagram
{}from the left.  Here $r$ is the number of pseudo-particles.
(b) $N_{n,m}$ which increases the number of pseudo-particle
{}from $m$ to $n$ without changing the Young diagram
associated with the state.
This operator can be realized as the
integral transformation for the wave function.
By those operators, the Jack polynomial for the
Young diagram $\lambda'=((r^1)^{s^1},(r^2)^{s^2},
\cdots,(r^{N-1})^{s^{N-1}})$
with $N_0$ variables is written
as
\be
  J_{\lambda}\propto N_{N_0,r^1}G_{s^1}
  N_{r^1,r^2}G_{s^2}\cdots
  N_{r^{N-2},r^{N-1}}G_{s^{N-1}}|\phi\rangle_{r^{N-1}}.
  \label{eAwata}
\ee
Here $|\phi\rangle_{r^{N-1}}$ is the vacuum for
$r^{N-1}$ pseudo-particles.
The following figure illustrates our construction
for $\beta=3$ and $r^{N-1}=3$ case in the momentum space
of the pseudo-particles.
Here $1$ 		
indicates 		
the momentum occupied by the pseudo-particles. 		

\vskip 4mm
\input epsf.tex
\centerline{\epsfbox{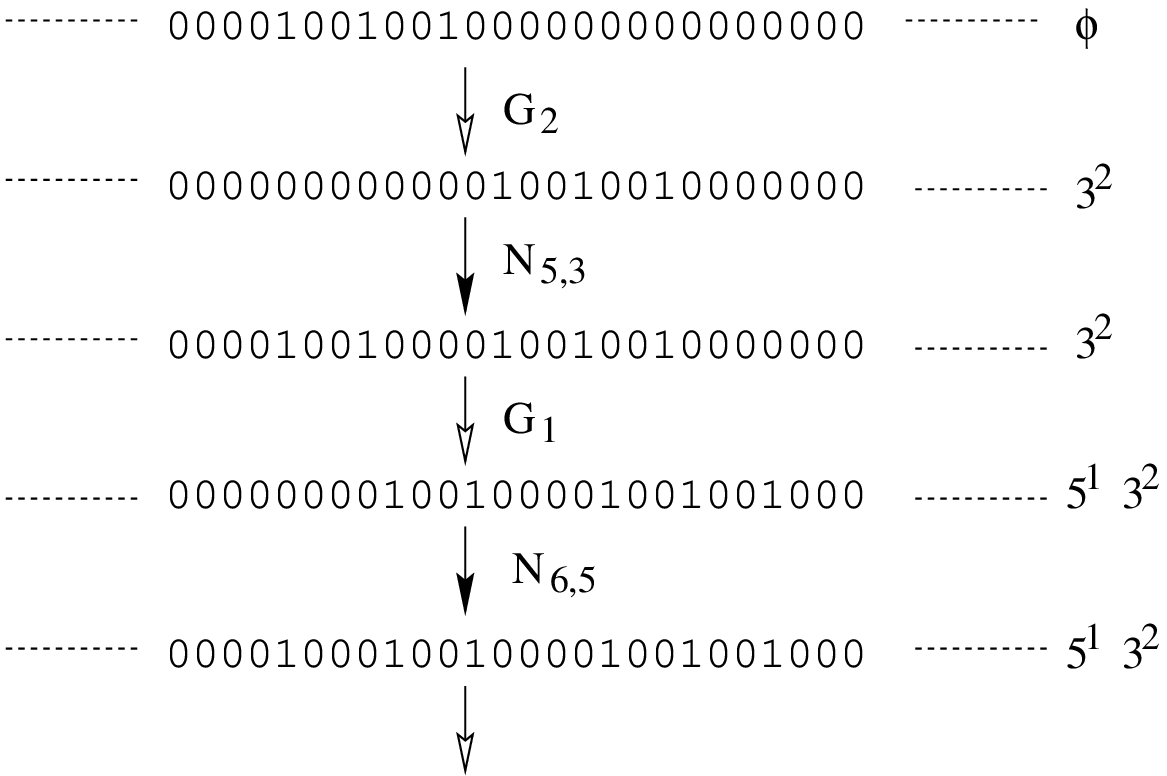}}
\centerline{{\bf Figure 1:} Construction of Jack Polynomials}
{~}\vspace{1mm}

This paper is organized as follows.
In section 2 we give a short summary of
the Calogero-Sutherland model and the Jack polynomial.
In section 3  we apply the collective field method to the
Calogero-Sutherland model. The Hamiltonian and momentum operators
are realized by the bosonic operators.
In section 4 we show the relation between the singular vectors of
the Virasoro algebra and the Jack symmetric
polynomials with the rectangular Young diagram.
This result is generalized in section 5.
The singular vectors of the $W_N$ algebra are related to the
Jack symmetric polynomials with the Young diagrams which
consist of $N-1$ rectangles.
Using this relations, we obtain the integral representations of
the Jack polynomial with arbitrary Young diagram.
In section 6, we define integral transformations
which directly give the Jack polynomials as we explain the formula \eq{eAwata}.
In section 7 we give the integral representations of the skew-Jack
polynomials.
Section 8 is devoted to discussions on many relevant topics.
In appendix A we discuss how the Jack
symmetric polynomials are realized on the boson Fock space.
In appendix B we discuss the analytic continuations of some integrals.
We give explicit examples in appendix C.

\section{Short summary of the Calogero-Sutherland model
and Jack polynomials}

The ground state of $H_{CS}$ is given by \cite{rCS}
\be
  \Delta_{CS}^{\beta}
  =
  \biggl(\frac{L}{\pi}\prod_{i,j=1 \atop i<j}^{N_0}
  \sin\frac{\pi}{L}(q_i-q_j)\biggr)^{\beta}
\ee
with the ground state energy
$E_0=\frac{1}{6}(\frac{\pi}{L})^2\beta^2(N_0^3-N_0)$.
{}For $\beta=1$, this is nothing but the free fermion
vacuum (Vandermonde determinant).
%

Let us make the coordinate transformation,
$x_j=e^{2\pi iq_j/L}$.
We are interested in the excited states of the form
$J_{\lambda}(x)\Delta_{CS}^{\beta}$, where $J_{\lambda}(x)$ is
the symmetric polynomial of the coordinates $x_i$.
Hamiltonian and momentum acted on $J_{\lambda}(x)$ are given by
\ba
  &&
  \Delta_{CS}^{-\beta}H_{CS}\Delta_{CS}^{\beta}
  =\frac{1}{2}\Bigl(\frac{2\pi}{L}\Bigr)^2 H_{\beta}+E_0, \qquad
  \Delta_{CS}^{-\beta}P_{CS}\Delta_{CS}^{\beta}
  =\frac{2\pi}{L}P, \n
  &&
  H_{\beta}=
  \sum_{i=1}^{N_0} D_i^2
  +\beta\sum_{i,j=1 \atop i<j}^{N_0}
  \frac{x_i+x_j}{x_i-x_j}(D_i-D_j),\quad
  P=\sum_{i=1}^{N_0}D_i,
\ea
where $D_i=x_i\frac{\partial}{\partial x_i}$.

Eigenfunctions of $H_{\beta}$ and $P$ are called in mathematical
literature as the Jack
symmetric polynomials \cite{rS}, $J_{\lambda}(x)$.
They are indexed by the Young diagram $\lambda$,
which may be physically interpreted as the distribution of
the momentum of pseudo-particles (holes) of the system.

The Young diagram is parametrized by the numbers of boxes in each
row, $\lambda=(\lambda_1,\cdots,\lambda_{N_0})$,
$\lambda_1\geq\cdots\geq\lambda_{N_0}\geq 0$.
The length $\ell(\lambda)$ of $\lambda$ is the number of the
non-zero $\lambda_i$'s.
Then $\lambda=(\lambda_1,\cdots,\lambda_{N_0})$ and
$(\lambda_1,\cdots,\lambda_{\ell(\lambda)})$ stand for the
same Young diagram.
The conjugate Young diagram is defined by interchanging rows with
columns, denoted by
$\lambda'=(\lambda'_1,\lambda'_2,\cdots)$ or
$(\lambda'_1,\cdots,\lambda'_{\lambda_1})$.
The total number of boxes is denoted by $|\lambda|=\sum_i\lambda_i$.

The energy eigenvalue was obtained as \cite{rCS},
\be
  \epsilon_{\beta,\lambda}
  =
  \sum_{i=1}^{N_0}\Bigl(\lambda_i^2+\beta(N_0+1-2i)\lambda_i\Bigr)
  =
  \sum_{i=1}^\infty 
\Bigl(-\beta\lambda_i^{\prime \, 2}
  +(\beta N_0+2i-1)\lambda_i'\Bigr).
  \label{ebl}
\ee
The eigenvalue of the momentum $P$ is $|\lambda|$.
Corresponding eigenvalues of $H_{CS}$ and $P_{CS}$ are
$\sum_{i=1}^{N_0}\frac{1}{2}k_i^2$ and $\sum_{i=1}^{N_0}k_i$,
respectively, where
\be
k_i=\frac{2\pi}{L}\Bigl(\lambda_i+\frac{\beta}{2}({N_0+1}-2i)\Bigr).
\ee
This formula gives the relation between the Young diagram and
the momentum distribution of pseudo-particles.
Since $\lambda_i$ is a decreasing 		
set of positive numbers,
there is a constraint for the neighboring
occupied momentum, $k_i-k_{i+1}\geq \beta\frac{2\pi}{L}$. This is
a realization of				
the generalized exclusion principle in the momentum space.

On the other hand, the second formula in \eq{ebl} shows that
the total energy is alternatively expressed as
$$constant - \frac{\beta}{2}\sum_{i\geq 1} \tilde{k}_i^2$$ where
\be
\tilde{k}_i=\frac{2\pi}{L}\Bigl(\lambda_i^\prime-\frac{1}{2\beta}(
\beta N_0-1+2i)\Bigr).
\ee
One may recognize that $\tilde{k}_i$'s are regarded as
the momenta of pseudo-holes. 
They are constrained by $\tilde{k}_i-\tilde{k}_{i+1}\geq \frac{2\pi}{\beta L}$.
By these observations, a Young diagram with $n$ rows (columns)
is regarded as describing
a state with a excitation of $n$ pseudo-particles (pseudo-holes).
Conjugating a Young diagram 		
is physically interpreted as
interchanging the pseudo-particles with parameter $\beta$ and pseudo-holes
with $1/\beta$.

In order to construct explicit form of the Jack polynomial,
it is important to understand the
mathematical structure of the Hilbert space.
It is identified with the ring of symmetric functions,
which has several basis, e.g., the power-sum
symmetric functions, the monomial symmetric functions and so on.
The power-sum symmetric function $p_{\lambda}(x)$ is defined by
$p_{\lambda}(x)=p_{\lambda_1}(x)\cdots p_{\lambda_M}(x)$, where
$p_n(x)=\sum_{i=1}^{N_0}x_i^n$.
The monomial symmetric function $m_{\lambda}(x)$ is defined by
$m_{\lambda}(x)=\sum_{\sigma}x_1^{\lambda_{\sigma(1)}}\cdots
x_{N_0}^{\lambda_{\sigma(N_0)}}$, where the summation is over
all distinct permutations of $(\lambda_1,\cdots,\lambda_{N_0})$.

The Jack symmetric polynomial
$J_{\lambda}(x)=J_{\lambda}(x;\beta)
=J_{\lambda}(x_1,\cdots,x_{N_0};\beta)$
is uniquely specified by the following two properties and normalization,
\ba
  \mbox{(\romannumeral1)}&&
  J_{\lambda}(x;\beta)
  =\sum_{\mu\leq\lambda}v_{\lambda,\mu}(\beta)m_{\mu}(x),\quad
  v_{\lambda,\lambda}(\beta)=1,\\
  \mbox{(\romannumeral2)}&&
  H_{\beta}J_{\lambda}(x;\beta)
  =\epsilon_{\beta,\lambda}J_{\lambda}(x;\beta).
\ea
In (i), we used
the dominance partial ordering on the Young diagrams defined as
$\lambda\geq\mu \Leftrightarrow |\lambda|=|\mu|$ and
$\lambda_1+\cdots+\lambda_i\geq\mu_1+\cdots+\mu_i$ for all $i$.


We introduce an inner-product on the Hilbert space
in the following manner \cite{rS},
\be
  \langle p_1^{k_1}\cdots p_n^{k_n},
  p_1^{\ell_1}\cdots p_m^{\ell_m}\rangle_{\beta}
  =
  \delta_{\vec k,\vec\ell}\,\beta^{-\sum_{i=1}^n k_i}
  \prod_{i=1}^n i^{k_i}k_i!,
  \label{ip}
\ee
for all $n,m\geq 1$.
This definition of inner-product is compatible with the
bosonization in the next section.
With this inner-product, the condition (\romannumeral2)
can be replaced by the orthogonality
condition,
\ba
  \mbox{(\romannumeral2)}'&&
  \langle J_{\lambda}(x;\beta),J_{\mu}(x;\beta)\rangle_{\beta}
  \propto\delta_{\lambda,\mu}.\hspace{25mm}
\ea
In section 6, we discuss another type of inner-product.

\section{Collective field method in the Calogero-Sutherland Model}

We will study $H_{\beta}$ by a collective field approach (bosonization).
Since $J_{\lambda}(x)$ is a symmetric function in $x_i$, it can
be written out using the power-sum polynomials $p_n$.
Therefore $H_{\beta}$ can be expressed in terms of creation and
annihilation of power-sums.
In conventional collective filed method,
power-sum appears as $p_n=\int dx \,x^n \rho(x)$,
where $\rho(x)$ is a density operator,
$\rho(x)=\sum_{i=1}^{N_0}\delta(x-x_i)$.

To realize creation and annihilation of power-sums,
we introduce a free boson field,
\ba
  &&
  \phi(z)=\hat{q}+a_0\log z-\sum_{n\neq 0}\sfrac{1}{n}a_nz^{-n},\quad
  \phi_-(z)=\sum_{n>0}\sfrac{1}{n}a_{-n}z^n, \n
  &&
  \lbrack a_n,a_m\rbrack=n\delta_{n+m,0},\quad
  \lbrack a_0,\hat{q}\rbrack=1.
  \label{phi}
\ea
Its operator product expansion is $\phi(z)\phi(w)\sim \log(z-w)$.
The normal ordering $:a_na_m:$ is defined by $a_na_m$ for $n\leq m$,
$a_ma_n$ for $n>m$ and $:\hat{q}a_0:\;=\;:a_0\hat{q}:=\hat{q}a_0$.
The boson Fock space $\cal{F}_{\alpha}$ is generated over oscillators
of negative mode by the state $\ket{\alpha}$ such that
\be
  a_n\ket{0}=0 \quad (n\geq 0),\qquad
  \ket{\alpha}=e^{\alpha\,\hat q}\ket{0}.
\ee
$\bra{\alpha}$ is similarly defined, with the normalization
$\bra{\alpha}\alpha'\rangle=\delta_{\alpha,\alpha'}$.

We consider the following map from a state $\ket{f}$ into
${\cal F}_{\alpha}$ to a symmetric function $f(x)$,
\ba
  \ket{f}\mapsto
  f(x)
  &\!\!=\!\!&
  \bra{\alpha}C_{\beta'}\ket{f}, \n
  C_{\beta'}
  &\!\!=\!\!&
  \exp\Bigl(\beta'\sum_{n>0}\sfrac{1}{n}a_np_n\Bigr),\quad
  p_n=\sum_ix_i^n,
  \label{sfc}
\ea
where $\beta'$ is a parameter.
Under this correspondence,
$a_{-n}$ and $a_n$ are interpreted as the creation and annihilation
operator of power-sum, $\beta'p_n$ and
$\frac{n}{\beta'}\frac{\partial}{\partial p_n}$,
respectively, because
$\bra{\alpha}C_{\beta'}a_{-n}=\beta'p_n\bra{\alpha}C_{\beta'}$ and
$\bra{\alpha}C_{\beta'}a_n=
\frac{n}{\beta'}\frac{\partial}{\partial p_n}\bra{\alpha}C_{\beta'}$.
We remark that 
after rescaling $a_n\rightarrow\sqrt{\beta}a_n$
the inner-product on the boson Fock space
agrees with that on the ring of symmetric functions \eq{ip}.

Hamiltonian and momentum can be expressed in terms of boson
oscillators as follows:
\be
  H_{\beta}\bra{\alpha}C_{\beta'}=\bra{\alpha}C_{\beta'}\hat{H}_{\beta},
  \quad
  P\bra{\alpha}C_{\beta'}=\bra{\alpha}C_{\beta'}\hat{P},
\ee
where
\be
  \hat{H}_{\beta}
  =
  \sum_{n,m>0}
  \Bigl(\beta'a_{-n-m}a_na_m
  +\frac{\beta}{\beta'}a_{-n}a_{-m}a_{n+m}\Bigr)
  +\sum_{n>0}a_{-n}a_n\Bigl((1-\beta)n+N_0\beta\Bigr),
\ee
and $\hat{P}=\sum_{n>0} a_{-n}a_n$.
We remark that $\hat{H}_{\beta}$ and $\hat{P}$ are independent
of $\alpha$.
Now the problem of finding the Jack polynomials is
translated to that of finding the eigenstates of
$\hat{H}_{\beta}$ and $\hat{P}$ in ${\cal F}_{\alpha}$.
In the rest of this section, the next section and appendix A,
we set $\beta'$ as
\be
  \sqrt{2}\beta'=\sqrt{\beta},
  \label{beta'2}
\ee
and define $\alpha_{\pm}$ as
\be
  \frac{\alpha_+}{\sqrt{2}}=\sqrt{\beta},\quad
  \frac{\alpha_-}{\sqrt{2}}=\frac{-1}{\sqrt{\beta}}.
\ee

The eigenstates for a single pseudo-particle (-hole) excitation,
or the Jack polynomials of the Young diagram with single row
(-column),
are expressed by a single vertex operator \cite{rS} in a boson language.
Its generating function is
\be
  e^{\alpha_{\pm}\phi_-(z)}
  =
  \sum_{n=0}^{\infty}\hat{J}^{\pm}_nz^n,
\ee
namely $\hat{J}^{\pm}_n\ket{\alpha}$ is the eigenstate
of $\hat{H}_{\beta}$ which corresponds to a Young diagram with
single row($+$) or single column($-$) with $n$ boxes, respectively.
By \eq{sfc}, we have
\be
  \bra{\alpha}C_{\beta'}e^{\alpha_{\pm}\phi_-(z)}\ket{\alpha}
  =
  \sum_{n=0}^{\infty}J_n^{\pm}(x)z^n
  =
  \prod_i(1-x_iz)^{-\beta'\alpha_{\pm}}.
\ee
In the appendix A, we discuss diagonalization of the Hamiltonian
by the operators $\hat J_n^\pm$.

\section{Virasoro singular vectors and Jack polynomials}

The Virasoro singular vector
is represented by the Jack polynomial with the rectangular
Young diagram \cite{rMY1}(see also \cite{rAMOS}).
In this section we give another proof, which will be
generalized in the next section.

Using a free boson \eq{phi}, the Virasoro algebra with the central
charge $c$ is realized as follows:
\ba
  T(z)
  &\!\!=\!\!&
  \sum_{n}L_nz^{-n-2}=
  \sfrac{1}{2}:\partial\phi(z)\partial\phi(z):
  +\alpha_0\partial^2\phi(z), \n
  c
  &\!\!=\!\!&
  1-12\alpha_0^2,\quad 2\alpha_0=\alpha_++\alpha_-.
\ea
The vertex operator $:e^{\alpha\phi(z)}:$ is a primary field
of the Virasoro algebra, and it creates the highest weight state
of the Virasoro algebra from the vacuum,
$\ket{\alpha}=\;:e^{\alpha\phi(0)}:\ket{0}$, whose conformal weight is
\be
  h(\alpha)=\frac{1}{2}\Bigl((\alpha-\alpha_0)^2-\alpha_0^2\Bigr).
\ee
We define $\alpha_{r,s}$ as
\be
  \alpha_{r,s}=\sfrac{1}{2}(1+r)\alpha_++\sfrac{1}{2}(1+s)\alpha_-
\ee
and remark that
\be
  h(\alpha_{\mp r,\pm s})=h(\alpha_{r,s})+rs.
\ee

In the Virasoro representation space with the highest weight state
$\ket{\alpha_{r,s}}$, we have a singular vector $\ket{\chi_{r,s}^+}$
at level $rs$. By using a screening current $:e^{\alpha_+\phi(z)}:$,
$\ket{\chi_{r,s}^+}$ is given as follows \cite{rKM}\cite{rTK}:
\ba
  \ket{\chi_{r,s}^+}
  &\!\!=\!\!&
  \oint\prod_{j=1}^r\frac{dz_j}{2\pi i}\cdot
  \prod_{i=1}^r:e^{\alpha_+\phi(z_i)}:
  \ket{\alpha_{-r,s}} \n
  &\!\!=\!\!&
  \oint\prod_{j=1}^r\frac{dz_j}{2\pi iz_j}\cdot
  \prod_{i,j=1 \atop i<j}^r(z_i-z_j)^{2\beta}\cdot
  \prod_{i=1}^rz_i^{(1-r)\beta-s}\cdot
  \prod_{j=1}^re^{\alpha_+\phi_-(z_j)}
  \ket{\alpha_{r,s}},
\ea
where the integration contour is shown in Figure 2(b),
which reduces to the contour in Figure 2(a) for a positive
integer $\beta$.
We note that this is just the form in \eq{JF}.
In \cite{rMY1}, they acted with $H_{\beta}$ on the polynomial
$\bra{\alpha_{r,s}}C_{\beta'}\ket{\chi_{r,s}^+}$ directly
and showed that it is an eigenfunction.
Here we will use $\hat{H}_{\beta}$.

{~}\vspace{3mm}
\centerline{\epsfbox{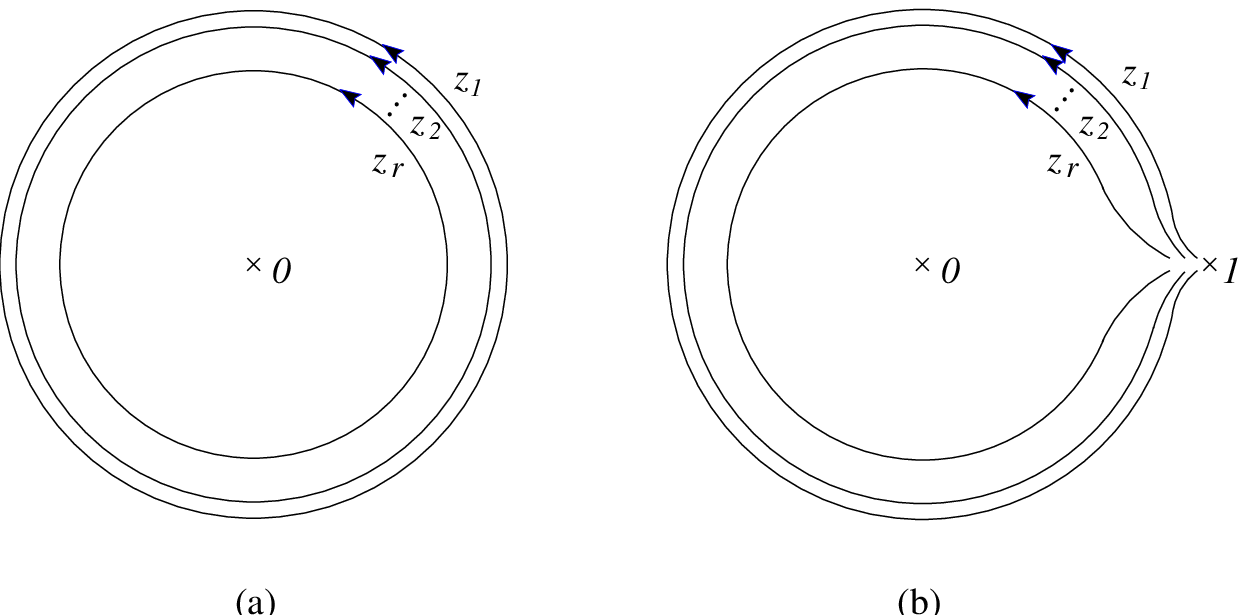}}
\centerline{{\bf Figure 2:} Integration contour}
{~}\vspace{1mm}

Under the choice of $\beta'$ \eq{beta'2}, the cubic term
in $H_{\beta}$ can be expressed by using the Virasoro generators.
This is the key point of our argument.
We have
%
\be
  \hat{H}_{\beta}
  =
  \sqrt{2\beta}\sum_{n>0}a_{-n}L_n
  +\sum_{n>0}a_{-n}a_n(N_0\beta+\beta-1-\sqrt{2\beta}a_0).
\ee
Since a singular vector $\ket{\chi_{r,s}^+}$ is annihilated
by $L_n$ ($n>0$), we have
\ba
  \hat{H}_{\beta}\ket{\chi_{r,s}^+}
  &\!\!=\!\!&
  \sum_{n>0}a_{-n}a_n(N_0\beta+\beta-1-\sqrt{2\beta}\alpha_{r,s})
  \ket{\chi_{r,s}^+} \n
  &\!\!=\!\!&
  rs\Bigl((N_0-r)\beta+s\Bigr)
  \ket{\chi_{r,s}^+}.
\ea
This eigenvalue is just $\epsilon_{\beta,\lambda}$ with the rectangular
Young diagram $\lambda=(s,s,\cdots,s)=(s^r)$.
Therefore
we obtain an integral representation of the Jack
polynomial with the Young diagram $\lambda=(s^r)$,
\ba
  {\cal N}_{r,s}^+ \,{\cal N}_{(s^r)}^+ J_{(s^r)}(x)
  &\!\!=\!\!&
  \bra{\alpha_{r,s}}C_{\beta'}\ket{\chi_{r,s}^+} \n
  &\!\!=\!\!&
  \oint\prod_{j=1}^r\frac{dz_j}{2\pi iz_j}\cdot
  \prod_{i,j=1 \atop i<j}^r(z_i-z_j)^{2\beta}\cdot
  \prod_{i=1}^rz_i^{(1-r)\beta-s}\cdot
  \prod_i\prod_{j=1}^r(1-x_iz_j)^{-\beta},
  \label{J+rs}
\ea
where the normalization constants ${\cal N}_\lambda^+$ \cite{rS}
and ${\cal N}_{r,s}^+$ (see appendix B) are given by
\be
  {\cal N}_{\lambda}^+
  =
  \prod_{s\in\lambda}
  \frac{(\ell_{\lambda}(s)+1)\beta+a_{\lambda}(s)}
       {\ell_{\lambda}(s)\beta+a_{\lambda}(s)+1}, \qquad
  {\cal N}_{r,s}^+
  =
  \frac{1}{r!}
  \prod_{j=1}^r\frac{\sin\pi j\beta}{\sin\pi\beta}\cdot
  \frac{\Gamma(r\beta+1)}{\Gamma(\beta+1)^r}.
  \label{Nrs}
\ee
Here for each box $s=(i,j)$ ($i$-th row and $j$-th column) in the
Young diagram $\lambda$, the arm-length $a_{\lambda}(s)$
and leg-length $\ell_{\lambda}(s)$ are defined by
$\lambda_i-j$ and $\lambda'_j-i$, respectively.
For a positive integer $\beta$, this ${\cal N}_{r,s}^+$ becomes
$(-1)^{\frac{1}{2}r(r-1)\beta}(r\beta)!/(\beta!)^r$.

There exists another screening current $:e^{\alpha_-\phi(z)}:$.
Using this, the singular vector is expressed in another form,
\be
  \ket{\chi_{r,s}^-}
  =
  \oint\prod_{j=1}^s\frac{dz_j}{2\pi i}\cdot
  \prod_{i=1}^s:e^{\alpha_-\phi(z_i)}:
  \ket{\alpha_{r,-s}},
\ee
which differs from $\ket{\chi_{r,s}^+}$ only normalization factor.
Similarly we can show that $\ket{\chi_{r,s}^-}$ is the eigenstate
of $\hat{H}_{\beta}$ and obtain
{
another integral representation of the Jack polynomial with the
Young diagram $(s^r)$},
\ba
  {\cal N}_{r,s}^- \,{\cal N}_{(r^s)}^- J_{(r^s)}(x)
  &\!\!=\!\!&
  \bra{\alpha_{r,s}}C_{\beta'}\ket{\chi_{r,s}^-} \n
  &\!\!=\!\!&
  \oint\prod_{j=1}^s\frac{dz_j}{2\pi iz_j}\cdot
  \prod_{i,j=1 \atop i<j}^s(z_i-z_j)^{2/\beta}\cdot
  \prod_{i=1}^sz_i^{(1-s)/\beta-r}\cdot
  \prod_i\prod_{j=1}^s(1-x_iz_j), \\
  {\cal N}_{\lambda}^-
  &\!\!=\!\!&
  (-1)^{|\lambda|},\qquad
  {\cal N}_{r,s}^-
  =
  \frac{1}{s!}
  \prod_{j=1}^s\frac{\sin\pi j\beta^{-1}}{\sin\pi\beta^{-1}}\cdot
  \frac{\Gamma(s\beta^{-1}+1)}{\Gamma(\beta^{-1}+1)^s}.
\ea

To illustlate the results obtained in this section,
we give explicit examples in appendix C.

\section{$W_N$ singular vectors and Jack polynomials}

\subsection{Review of $W_N$ algebra}
To discuss the $W_N$ algebra, we start fixing our notation for
$A_{N-1}$. Let $\vec{e}_i$ ($i=1,\cdots,N$) to be
an orthonormal basis ($\vec{e}_i\cdot\vec{e}_j=\delta_{ij}$), and
the weight space of $A_{N-1}$ to be the hyper-surface perpendicular to
$\sum_{i=1}^N\vec{e}_i$.
The weights of the vector representation $\vec{h}_i$ ($i=1,\cdots,N$),
the simple roots $\vec{\alpha}^a$ $(a=1,\cdots,N-1$), and
the fundamental weights $\vec{\Lambda}_a$ $(a=1,\cdots,N-1$),
are given by
\be
  \vec{h}_i=\vec{e}_i-\sfrac{1}{N}\sum_{j=1}^N\vec{e}_j,\quad
  \vec{\alpha}^a=\vec{h}_a-\vec{h}_{a+1},\quad
  \vec{\Lambda}_a=\sum_{i=1}^a\vec{h}_i,
\ee
and their inner-products are
\ba
  \vec{\alpha}^a\cdot\vec{\alpha}^b
  &\!\!=\!\!&
  A^{ab}=2\delta^{a,b}-\delta^{a,b+1}-\delta^{a,b-1}, \n
  \vec{\alpha}^a\cdot\vec{\Lambda}_b
  &\!\!=\!\!&
  A^a_b=\delta^{a,b}, \n
  \vec{\Lambda}_a\cdot\vec{\Lambda}_b
  &\!\!=\!\!&
  A^{-1}_{ab}=\sfrac{1}{N}\min(a,b)\Bigl(N-\max(a,b)\Bigr).
\ea
Components of a vector $\vec{X}$ in the weight space are defined by
\be
  \vec{X}
  =\sum_{a=1}^{N-1}X^a\vec{\Lambda}_a
  =\sum_{a=1}^{N-1}X_a\vec{\alpha}^a.
\ee
They are related each other,
$X_a=\sum_bA^{-1}_{ab}X^b$, $X^a=\sum_bA^{ab}X_b$
and we use the convention $X_0=X_N=X^0=X^N=0$.
{}From an orthonormal boson
$\vec{\varphi}(z)=\sum_{i=1}^N\varphi_i(z)\vec{e}_i$
($\varphi_i(z)\varphi_j(w)\sim\delta_{ij}\log(z-w)$),
we define
$\vec{\phi}(z)=\vec{\varphi}(z)
-(\vec{\varphi}(z)\cdot\sum_{j=1}^N\vec{e}_j)
\frac{1}{N}\sum_{i=1}^N\vec{e}_i
=\sum_{a=1}^{N-1}\phi^a(z)\vec{\Lambda}_a
=\sum_{a=1}^{N-1}\phi_a(z)\vec{\alpha}^a$,
namely,
\ba
  &&
  \vec{\phi}(z)=\vec{\hat{q}}+\vec{a}_0\log z
  -\sum_{n\neq 0}\sfrac{1}{n}\vec{a}_nz^{-n},\quad
  \vec{\phi}_-(z)=\sum_{n>0}\sfrac{1}{n}\vec{a}_{-n}z^n, \n
  &&
  \lbrack a_n^a,a_m^b\rbrack=A^{ab}n\delta_{n+m,0},\quad
  \lbrack a_0^a,\hat{q}^b\rbrack=A^{ab},
\ea
with operator product expansion
$\phi^a(z)\phi^b(w)\sim A^{ab}\log(z-w)$.
The boson Fock space $\cal{F}_{\vec{\lambda}}$ is generated
by oscillators of negative mode on the state $\ket{\vec{\lambda}}$,
which is characterized by
\be
  \vec{a}_n\ket{\vec{0}}=0 \quad (n\geq 0),\qquad
  \ket{\vec{\lambda}}=e^{\vec{\lambda}\cdot\vec{\hat q}} \ket{\vec{0}}.
\ee
$\bra{\vec{\lambda}}$ is similarly defined, with the normalization
$\bra{\vec{\lambda}}\vec{\lambda}'\rangle=
\delta_{\vec{\lambda},\vec{\lambda}'}$.


Generators of the $W_N$ algebra, $W^k(z)$ ($k=2,\cdots,N$), are
obtained by the Miura transformation \cite{rFL},
\ba
  &&
  :\prod_{i=1}^N\Bigl(
  \alpha_0\partial+\vec{h}_i\cdot\partial\vec{\phi}(z)\Bigr):\;
  =
  \sum_{k=0}^NW^k(z)(\alpha_0\partial)^{N-k}, \n
  &&
  \alpha_0=\alpha_++\alpha_-, \quad
  \alpha_+=\sqrt{\beta},\quad \alpha_-=\frac{-1}{\sqrt{\beta}}.
\ea
{}From this, the Virasoro generator with the central charge $c$ is
given by
\ba
  T(z)
  &\!\!=\!\!&
  -W^2(z)=
  \sfrac{1}{2}:\partial\vec{\phi}(z)\cdot\partial\vec{\phi}(z):
  +\alpha_0\vec{\rho}\cdot\partial^2\vec{\phi}(z), \n
  c
  &\!\!=\!\!&
  N-1-12\alpha_0^2\vec{\rho}^{\,2},
\ea
where
$\vec{\rho}$ is the half-sum of positive roots,
$\vec{\rho}=\sum_{a=1}^{N-1}\vec{\Lambda}_a$, and
$\vec{\rho}^{\,2}=\frac{1}{12}N(N^2-1)$.
The $W^3$ generator $W(z)$ is given by
\ba
  W(z)
  &\!\!=\!\!&
  W^3(z)=\sum_nW_nz^{-n-3} \n
  &\!\!=\!\!&
  \sum_{a=1}^{N-1}:\partial\phi_a(z)\partial\phi_a(z)
  \Bigl(\partial\phi_{a+1}(z)-\partial\phi_{a-1}(z)\Bigr): \n
  &&
  +\alpha_0\sum_{a,b=1}^{N-1}(1-a)A^{ab}
  :\partial\phi_a(z)\partial^2\phi_b(z):
  +\alpha_0^2\sum_{a=1}^{N-1}(1-a)\partial^3\phi_a(z).
\ea
The vertex operator $:e^{\vec{\lambda}\cdot\vec{\phi}(z)}:$ is
a primary field of the $W_N$ algebra and creates the highest weight
state of the $W_N$ algebra from the vacuum,
$\ket{\vec{\lambda}}=
\;:e^{\vec{\lambda}\cdot\vec{\phi}(0)}:\ket{\vec{0}}$.
Its conformal weight $h(\vec{\lambda})$ and $W_0$-eigenvalue
$w(\vec{\lambda})$ are
\ba
  h(\vec{\lambda})
  &\!\!=\!\!&
  \sfrac{1}{2}\Bigl((\vec{\lambda}-\alpha_0\vec{\rho})^2
  -\alpha_0^2\vec{\rho}^{\,2}\Bigr), \n
  w(\vec{\lambda})
  &\!\!=\!\!&
  \sum_{a=1}^{N-1}
  \biggl(\lambda_a\lambda_a(\lambda_{a+1}-\lambda_{a-1}) \n
  &&
  \qquad
  +\alpha_0\Bigl(2(a-1)\lambda_a\lambda_a
  +(1-2a)\lambda_a\lambda_{a+1}\Bigr)
  +\alpha_0^22(1-a)\lambda_a\biggr).
\ea
We define $\vec{\lambda}_{\vec{r},\vec{s}}^{\pm}$ as follows\footnote{
Usual parametrization of the weight vector is
$\vec{\lambda}_{\vec{r},\vec{s}}
=\sum_{a=1}^{N-1}((1+r^a)\alpha_++(1+s^a)\alpha_-)\vec{\Lambda}_a
=\alpha_+\vec{r}+\alpha_-\vec{s}+\alpha_0\vec{\rho}$.
}
\ba
  \vec{\lambda}_{\vec{r},\vec{s}}^+
  &\!\!=\!\!&
  \sum_{a=1}^{N-1}\Bigl((1+r^a-r^{a-1})\alpha_++(1+s^a)\alpha_-\Bigr)
  \vec{\Lambda}_a, \\
  \vec{\lambda}_{\vec{r},\vec{s}}^-
  &\!\!=\!\!&
  \sum_{a=1}^{N-1}\Bigl((1+r^a)\alpha_++(1+s^a-s^{a-1})\alpha_-\Bigr)
  \vec{\Lambda}_a.
\ea
We remark that
\be
  h\Bigl(\vec{\lambda}_{\vec{r},\vec{s}}^{\pm}
  -\alpha_{\pm}\sum_{a=1}^{N-1}r^a_{\pm}\vec{\alpha}^a\Bigr)
  =
  h(\vec{\lambda}_{\vec{r},\vec{s}}^{\pm})
  +\sum_{a=1}^{N-1}r^as^a,
\ee
where $r^a_+=r^a$ and $r^a_-=s^a$.

A singular vector $\ket{\chi_{\vec{r},\vec{s}}^+}$ at
level $\sum_{a=1}^{N-1}r^as^a$ in the $W_N$ representation
space with the highest weight state
$\ket{\vec{\lambda}_{\vec{r},\vec{s}}^+}$ is expressed by
using screening currents $:e^{\alpha_+\phi^a(z)}:$,
\ba
  \ket{\chi_{\vec{r},\vec{s}}^+}
  &\!\!=\!\!&
  \oint\prod_{a=1}^{N-1}\prod_{j=1}^{r^a}\frac{dz^a_j}{2\pi i}\cdot
  \prod_{a=1}^{N-1}\prod_{j=1}^{r^a}:e^{\alpha_+\phi^a(z^a_j)}:
  \ket{\vec{\lambda}_{\vec{r},\vec{s}}^+
  -\alpha_+\sum_{a=1}^{N-1}r^a\vec{\alpha}^a} \n
  &\!\!=\!\!&
  \oint\prod_{a=1}^{N-1}\prod_{j=1}^{r^a}\frac{dz^a_j}{2\pi iz^a_j}\cdot
  \prod_{a=1}^{N-1}\prod_{i,j=1 \atop i<j}^{r^a}
  (z^a_i-z^a_j)^{2\beta}\cdot
  \prod_{a=1}^{N-2}\prod_{i=1}^{r^a}\prod_{j=1}^{r^{a+1}}
  (z^a_i-z^{a+1}_j)^{-\beta} \n
  &&
  \times\prod_{a=1}^{N-1}\prod_{j=1}^{r^a}
  (z^a_j)^{(1-r^a+r^{a+1})\beta-s^a}\cdot
  \prod_{a=1}^{N-1}\prod_{j=1}^{r^a}e^{\alpha_+\phi^a_-(z^a_j)}
  \ket{\vec{\lambda}_{\vec{r},\vec{s}}^+},
\ea
where we use a similar integration contour in Figure 2(b).
In the following we assume $r^1>\cdots>r^{N-1}$.

Similarly, using another screening currents $:e^{\alpha_-\phi^a(z)}:$,
we have a singular vector,
\be
  \ket{\chi_{\vec{r},\vec{s}}^-}
  =
  \oint\prod_{a=1}^{N-1}\prod_{j=1}^{s^a}\frac{dz^a_j}{2\pi i}\cdot
  \prod_{a=1}^{N-1}\prod_{j=1}^{s^a}:e^{\alpha_-\phi^a(z^a_j)}:
  \ket{\vec{\lambda}_{\vec{r},\vec{s}}^-
  -\alpha_-\sum_{a=1}^{N-1}s^a\vec{\alpha}^a},
\ee
with $s^1>\cdots>s^{N-1}$.

\subsection{$W_N$ Singular Vectors and Jack polynomials}
Like as \eq{sfc}, we consider the map from a state $\ket{f}$ in
${\cal F}_{\vec{\lambda}}$ into a symmetric function $f(x)$.
Since the Hilbert space of the Calogero-Sutherland model
is equivalent to a single boson Fock space, we have to reduce the
degree of freedoms  of the $N-1$ boson Fock space of $W_N$ algebra.
To this end, we use a kind of projection to give
the correspondence between the Hilbert spaces,
\ba
  f(x)
  &\!\!=\!\!&
  \bra{\vec{\lambda}}C_{\beta'}\ket{f}, \n
  C_{\beta'}
  &\!\!=\!\!&
  \exp\Bigl(\vec{\beta}'\cdot\sum_{n>0}\sfrac{1}{n}\vec{a}_np_n\Bigr),
  \quad
  \vec{\beta}'=\beta'\vec{\Lambda}_1,\quad
  p_n=\sum_ix_i^n,
  \label{sfcN}
\ea
where $\beta'$ is a parameter.
In this definition, only $a^1_{-n}$ is a creation operator of $p_n$ and
other $a^a_{-n}$ ($a>1$, $n>1$) do not contribute this map
because $\bra{\vec{\lambda}}C_{\beta'}a^a_{-n}
=\bra{\vec{\lambda}}C_{\beta'}\delta^{a1}\beta'p_n$.
The Hamiltonian and momentum can be expressed in terms of boson
oscillators as follows:
\be
  H_{\beta}\bra{\vec{\lambda}}C_{\beta'}
  =\bra{\vec{\lambda}}C_{\beta'}\hat{H}_{\beta},
  \quad
  P\bra{\vec{\lambda}}C_{\beta'}
  =\bra{\vec{\lambda}}C_{\beta'}\hat{P},
\ee
where
\be
  \hat{H}_{\beta}
  =
  \sum_{n,m>0}
  \Bigl(\beta'a^1_{-n-m}a_{1,n}a_{1,m}
  +\frac{\beta}{\beta'}a^1_{-n}a^1_{-m}a_{1,n+m}\Bigr)
  +\sum_{n>0}a^1_{-n}a_{1,n}\Bigl((1-\beta)n+N_0\beta\Bigr),
\ee
and $\hat{P}=\sum_{n>0} a^1_{-n}a_{1,n}$.
Since the above map is not one to one, $\hat{H}_{\beta}$ and $\hat{P}$
are determined up to the term
$\sum_{n>0}\sum_{a>1}a^a_{-n}\times(\cdots)$.
In the following we set $\beta'$ as
\be
  \beta'=\sqrt{\beta}.
  \label{beta'N}
\ee

Like as the Virasoro case, a straightforward calculation shows that
under the choice of $\beta'$ \eq{beta'N}, cubic term in
$\hat{H}_{\beta}$ can be expressed by the Virasoro and $W$ generators.
We have
\ba
  \hat{H}_{\beta}
  &\!\!=\!\!&
  \frac{2}{N}\sqrt{\beta}\sum_{n>0}a^1_{-n}L_n
  +\sum_{n>0}\vec{a}_{-n}\cdot\vec{a}_n
  (N_0\beta+\beta-1-2\sqrt{\beta}a_{1,0}) \n
  &&
  +\sqrt{\beta}(W_0-W_{0,zero})
  +\sum_{n>0}\sum_{a>1}a^a_{-n}\times(\cdots).
\ea
Here $W_{0,zero}$ is the zero mode part of $W_0$,
\ba
  W_{0,zero}
  &\!\!=\!\!&
  \sum_{a=1}^{N-1}\biggl(
  a_{a,0}a_{a,0}(a_{a+1,0}-a_{a-1,0}) \n
  &&\quad
  +\alpha_0\Bigl(2(a-1)a_{a,0}a_{a,0}+(1-2a)a_{a,0}a_{a+1,0}\Bigr)
  +\alpha_0^22(1-a)a_{a,0}\biggr).
\ea
Using these, we have
\ba
  \hat{H}_{\beta}\ket{\chi_{\vec{r},\vec{s}}^+}
  &\!\!=\!\!&
  \biggl(\sum_{a=1}^{N-1}r^as^a\times
  \Bigl(N_0\beta+\beta-1-2\sqrt{\beta}
  (\alpha_+r_1+\alpha_-s_1+\alpha_0\rho_1)\Bigr) \n
  &&
  \quad +\sqrt{\beta}\Bigl(
  w(\vec{\lambda}_{\vec{r},\vec{s}}^+
    -\alpha_+\sum_{a=1}^{N-1}r^a\vec{\alpha}^a)
  -w(\vec{\lambda}_{\vec{r},\vec{s}}^+)\Bigr)
  \biggr)
  \ket{\chi_{\vec{r},\vec{s}}^+} \n
  &&
  +\sum_{n>0}\sum_{a>1}a^a_{-n}\times(\cdots)
  \ket{\chi_{\vec{r},\vec{s}}^+} \n
  &\!\!=\!\!&
  \biggl(\sum_{a=1}^{N-1}r^as^as^a
  +2\sum_{a,b=1 \atop a>b}^{N-1}r^as^as^b
  +\beta\sum_{a=1}^{N-1}r^as^a(N_0-r^a)\biggr)
  \ket{\chi_{\vec{r},\vec{s}}^+} \n
  &&
  +\sum_{n>0}\sum_{a>1}a^a_{-n}\times(\cdots)
  \ket{\chi_{\vec{r},\vec{s}}^+}.
\ea
Therefore $\ket{\chi_{\vec{r},\vec{s}}^+}$ is the eigenstate
of $\hat{H}_{\beta}$ up to the last term, which will vanish
after multiplying by 		
$\bra{\vec{\lambda}_{\vec{r},\vec{s}}^+}C_{\beta'}$.
This eigenvalue is just $\epsilon_{\beta,\lambda}$ with
$\lambda'=((r^1)^{s^1},(r^2)^{s^2},\cdots,(r^{N-1})^{s^{N-1}})$,
namely,
$\lambda=((s^1+\cdots+s^{N-1})^{r^{N-1}},
(s^1+\cdots+s^{N-2})^{r^{N-2}-r^{N-1}},\cdots,(s^1)^{r^1-r^2})$,

\generalYoung

\noindent
Using the state--function correspondence \eq{sfcN}, we obtain
%
{
an integral representation of the Jack polynomial with the
Young diagram $\lambda$,
\ba
  J_{\lambda}(x)
  &\!\!=\!\!&
  \bigl({\cal N}_{\vec{r},\vec{s}}^+ \,{\cal N}_\lambda^+\bigr)^{-1}
  \bra{\vec{\lambda}_{\vec{r},\vec{s}}^+}C_{\beta'}
  \ket{\chi_{\vec{r},\vec{s}}^+} \n
  &\!\!=\!\!&
  \bigl({\cal N}_{\vec{r},\vec{s}}^+ \,{\cal N}_\lambda^+\bigr)^{-1}
  \oint\prod_{a=1}^{N-1}\prod_{j=1}^{r^a}\frac{dz^a_j}{2\pi iz^a_j}\cdot
  \prod_{a=1}^{N-1}\prod_{i,j=1 \atop i<j}^{r^a}
  (z^a_i-z^a_j)^{2\beta}\cdot
  \prod_{a=1}^{N-2}\prod_{i=1}^{r^a}\prod_{j=1}^{r^{a+1}}
  (z^a_i-z^{a+1}_j)^{-\beta} \n
  &&
  \hspace{25mm}
  \times\prod_{a=1}^{N-1}\prod_{j=1}^{r^a}
  (z^a_j)^{(1-r^a+r^{a+1})\beta-s^a}\cdot
  \prod_i\prod_{j=1}^{r^1}(1-x_iz^1_j)^{-\beta},
  \label{J+vrvs}
\ea
where normalization constants
${\cal N}_\lambda^+$ and ${\cal N}_{\vec{r},\vec{s}}^+$
are in \eq{Nrs} and appendix B, respectively.

Similarly, using another screening currents $:e^{\alpha_-\phi^a(z)}:$,
we obtain
{
another integral representation of the Jack polynomial with
the Young diagram
$\lambda=((s^1)^{r^1},(s^2)^{r^2},\cdots,(s^{N-1})^{r^{N-1}})$,}
\ba
  J_{\lambda'}(x)
  &\!\!=\!\!&
  \bigl({\cal N}_{\vec{r},\vec{s}}^- \,{\cal N}_\lambda^-\bigr)^{-1}
  \bra{\vec{\lambda}_{\vec{r},\vec{s}}^-}C_{\beta'}
  \ket{\chi_{\vec{r},\vec{s}}^-} \n
  &\!\!=\!\!&
  \bigl({\cal N}_{\vec{r},\vec{s}}^- \,{\cal N}_\lambda^-\bigr)^{-1}
  \oint\prod_{a=1}^{N-1}\prod_{j=1}^{s^a}\frac{dz^a_j}{2\pi iz^a_j}\cdot
  \prod_{a=1}^{N-1}\prod_{i,j=1 \atop i<j}^{s^a}
  (z^a_i-z^a_j)^{2/\beta}\cdot
  \prod_{a=1}^{N-2}\prod_{i=1}^{s^a}\prod_{j=1}^{s^{a+1}}
  (z^a_i-z^{a+1}_j)^{-1/\beta} \n
  &&
  \hspace{25mm}
  \times\prod_{a=1}^{N-1}\prod_{j=1}^{s^a}
  (z^a_j)^{(1-s^a+s^{a+1})/\beta-r^a}\cdot
  \prod_i\prod_{j=1}^{s^1}(1-x_iz^1_j).
  \label{J-vrvs}
\ea


\section{Direct derivation of the integral formulas}

\subsection{Integral transformations}
In this section, we describe a direct method
to give the integral representation of the Jack polynomial.
We hope that it may give some insight
on the symmetry structure of the Calogero-Sutherland model.

Our method is based on two types of transformations
which map any eigenstates of the Hamiltonian into another.
The first transformation may be physically interpreted as
a global Galilean transformation which describes a uniform
shift of momentum of the pseudo-particles.
The second one is defined as the integral transformation
which changes the number of particles without touching
the Young diagram of the original Jack polynomial.

The first one, the Galilean transformation $G_s$,
is defined by,
\be
(G_s \psi) (x_1,\cdots,x_r)
= \prod_{i=1}^r (x_i)^s\cdot \psi(x_1,\cdots,x_r),
\ee
for any symmetric function $\psi$.
Recalling the definition of $x_j = e^{i\frac{2\pi}{L}q_j}$, it produces
a uniform shift of the momentum of the pseudo-particles,
\be
k_i \rightarrow k_i+\frac{2\pi}{L} s.
\ee
Therefore, when it operates on the Jack polynomial,
$G_s$ adds a rectangle Young diagram $(s^r)$ to
the original one from the left,
\be
G_s J_\lambda(x;\beta) = J_{\lambda+(s^r)}(x;\beta),
\label{e6.5}
\ee
$$
\Galilei
$$
By the definition of the Jack polynomial,
the normalization factor is one.

To define the second (integral) transformation,
we prepare some notations.
Let ${\x a{ }}\equiv ({\x a1},\cdots,{\x a{r^a}})$, $a\in\bZ_{\geq 0}$
be finite or infinite sequences of independent variables and denote
\ba
{\Daa { }}&\equiv&\prod_{i,j=1 \atop i\neq j}^{r^{ }}
	\left(1-{{\x { }i}/{\x { }j}}\right)
	\propto \Delta_{CS}(x)^2,\n
{\Dab ab}&\equiv&\prod_{i=1}^{r^a} \prod_{j=1}^{r^b}
\left(1-{{\x ai}/{\x bj}}\right).
\label{e6.1}
\ea
%
%
There are two types of the inner-products between
the symmetric polynomial with which the Jack
polynomial becomes mutually orthogonal.
The first one, $\langle\;,\;\rangle_\beta$, is defined in \eq{ip}.
%
The second one $\langle\;,\;{\sRangle\beta r}$ as \cite{rMac}
is defined for a positive integer $\beta$ by
\be
\langle f(x),g(x){\sRangle\beta r}
= {\Sca xjr} \cdot f(1/x)\, g(x) {\Daa { }}^\beta.
\label{e6.2}
\ee
Here the integral $\oint{\res xjr} f(x)$
stands for a constant part of $f(x)$.
%
The second one appears in the computation
of dynamical correlation functions \cite{rSLA}--\cite{rMP2}.
These two definitions are equivalent only when $\beta=1$.
The norm of the Jack polynomial is related by \cite{rMac},
for a positive integer $\beta$,
\ba
\langle J_\lambda, J_\lambda {\sRangle\beta r}
&\!\!\!\!\!\!\!\!&
= \prod_{1\leq i<j \leq r} \prod_{k=1}^{\beta-1}
{ \lambda_i - \lambda_j + k + \beta(j-i) \over
  \lambda_i - \lambda_j - k + \beta(j-i) },\n
&\!\!\!\!\!\!\!\!&
= \prod_{i=2}^r \left( { i\beta-1 \atop \beta-1 }\right)\cdot
\prod_{(i,j)\in\lambda} { j-1 + \beta(r-i+1) \over j + \beta(r-i) }\cdot
\langle J_\lambda, J_\lambda\rangle_\beta,
\label{e6.11}
\ea
where $r$ is a number of variables,
$\lambda=(\lambda_1\geq\cdots\geq\lambda_r\geq 0)$ and
$(i,j)\in\lambda$ is a square in the Young diagram such that
$1\leq i\leq \lambda'_1$ and $1\leq j\leq \lambda_i$.
Note that
$
\langle 1,1{\sRangle\beta r} =
\prod_{i=2}^r \left( { i\beta-1 \atop \beta-1 }\right) =
{ (r\beta)!/ r!(\beta!)^r}
$.


After this preparation, we introduce the
second integral transformation as
\be
(N^{(\beta)}_{r^a,r^b}\psi)(x^a_1,\cdots,x^a_{r^a})=
\oint{\res{x^b}j{r^b}}\cdot
{\Dab ab}^{-\beta} {\Daa b}^\beta
\psi(x^b_1,\cdots,x^b_{r^b}).
\label{IntTrans}
\ee
It transforms any eigenstate into another by
the orthogonality relations of the Jack polynomials
\ba
&\!\!\!\!\!\!\!\!&
{\Dab ab}^{-\beta}
=\sum_\lambda J_\lambda({\x a{ }};\beta) \,
J_\lambda\left({1/{\x b{ }}};\beta\right)
\langle J_\lambda,J_\lambda\rangle_\beta^{-1} ,\n
&\!\!\!\!\!\!\!\!&
{\Sca xjr}\cdot
J_\lambda(x;\beta) J_\mu\left({1/x};\beta\right) {\Daa { }}^\beta
= \delta_{\lambda,\mu} \langle J_\lambda,J_\lambda{\sRangle\beta r}.
\label{e6.3}
\ea
With  the normalization factor, we get,
\be
J_\lambda({\x a{ }};\beta) =
{    \langle J_\lambda,J_\lambda\rangle_\beta \over
r^b! \langle J_\lambda,J_\lambda{\sRangle\beta{r^b}} }
N^{(\beta)}_{r^a,r^b}J_\lambda({\x b{ }};\beta).
\label{e6.4}
\ee
As we discussed, it changes
the number of pseudo-particles without touching the Young diagram.
%
%

These two transformations
are enough to give the Jack polynomials with arbitrary Young
diagrams \cite{rMY2}.  Namely, by the Galilean transformation,
we can add a rectangle to the arbitrary Young diagram.
The difficulty was that
the number of rows of such rectangle is constrained
by the number of pseudo-particle.  However, we may change
it by the second transformation.
Any Jack polynomial can be constructed from the trivial
state, the vacuum, by the iterative use of
them. We arrive at
the integral representation of the Jack polynomials,
\ba
J_\lambda ({\x 0{ }};\beta)
&\!\!\!\!\!\!\!\!&
= N^{(\beta)}_{r^0,r^1} G_{s^1} N^{(\beta)}_{r^1,r^2}\cdots\cdots
  G_{s^{N-2}}N^{(\beta)}_{r^{N-2},r^{N-1}}G_{s^{N-1}}\cdot 1 \n
&\!\!\!\!\!\!\!\!&
= C^+_\lambda \oint\prod_{a=1}^{N-1} {\res{x^a}j{r^a}}\cdot
\prod_{a=1}^{N-1}{\Dab {a-1}a}^{-\beta} {\Daa a}^\beta 
\prod_{j=1}^{r^a}\left({\x aj}\right)^{s^a},
\label{e6.6}
\ea
\be
C^+_\lambda
=\prod_{a=1}^{N-1}
{   \langle J_{{\laa a}},J_{{\laa a}}\rangle_\beta  \over
r^a!\langle J_{{\laa a}},J_{{\laa a}}{\sRangle\beta{r^a}}  },
\label{C+}
\ee
with
${\laa a}'\equiv((r^a)^{s^a},(r^{a+1})^{s^{a+1}},\cdots,(r^{N-1})^{s^{N-1}})$
and $\lambda\equiv\lambda^{(1)}$.
If we replace respectively ${\x 0i}$ and ${\x ai}$ $(a\neq 0)$
with $x_i$ and $1/z^a_i$,
this formula \eq{e6.6} reduces to the integral formula \eq{J+vrvs}.
The relation between the normalization constant $C^+_\lambda$
and that of the previous sections is
$(-1)^{\sum_a(r^a+\frac{1}{2}r^a(r^a-1)\beta)}C^+_\lambda =
 ({\cal N}^+_\lambda {\cal N}^+_{\vec{r},\vec{s}})^{-1}$.

\subsection{Dual transformations}
We next consider the dual orthogonal relation.
It is defined by the automorphism $\omega_\beta$ \cite{rS},
\be
\omega_\beta\,p_n \equiv (-1)^{n-1} \beta^{-1} p_n, .
\ee
for $n\neq 0$.
It satisfies
\ba
\omega_\beta^{(a)} {\Dab ab}^{-\beta}
&\!\!\!\!\!\!\!\!&
= {\tDab ab}
\equiv\prod_{i,j} \left(1 + {{\x ai}/{\x bj}}\right),\n
\omega_\beta\,J_\lambda(x;\beta)
\langle J_\lambda,J_\lambda\rangle_\beta^{-1}
&\!\!\!\!\!\!\!\!&
= J_{\lambda'}(x;1/\beta) ,
\label{e6.7}
\ea
where $\omega_{\beta}^{(a)}$ is acted on the variables ${\x ai}$'s.
The second relation shows that it interchanges rows
and columns of the Young diagram.  Physically, it amounts to interchange
pseudo-particles and pseudo-holes with the change
of the parameter $\beta \leftrightarrow 1/\beta$.
Some aspects of this transformation was discussed in
 \cite{rS} and \cite{rMP2}.

%
%
Using this automorphism $\omega_{\beta}^{(a)}$,
we get the following dual orthogonality relation,
\be
{\tDab ab}
=\sum_\lambda J_{\lambda'}\left({\x a{ }};1/\beta\right)
J_\lambda(1/{\x b{ }};\beta).
\label{e6.8}
\ee

We can introduce an integral transformation
which realizes the duality,
\be
(\widetilde N^{(\beta)}_{r^a,r^b} \psi)({\x a{ }})
\equiv \oint{\res{x^b}j{r^b}}\cdot{\tDab ab}{\Daa b}^\beta
\psi({\x b{ }}).
\ee
It maps into
\be
J_{\lambda'}({\x a{ }};1/\beta) =
{ 1 \over r^b! \langle J_\lambda,J_\lambda{\sRangle\beta{r^b}} }
\widetilde N^{(\beta)}_{r^a,r^b}
J_\lambda({\x b{ }};\beta).
\label{e6.9}
\ee
%
%
Applying the automorphism $\omega_{\beta}^{(0)}$
to the eq.\ \eq{e6.6} and replacing $\beta$ with $1/\beta$,
we obtain the dual form of the Jack polynomials:
\ba
&\!\!\!\!\!\!\!\!&
J_{\lambda'} ({\x 0{ }};\beta)
= C^-_\lambda\widetilde N^{(1/\beta)}_{r^0,r^1}
G_{s^1} N^{(1/\beta)}_{r^1,r^2}\cdots
  G_{s^{N-2}}N^{(1/\beta)}_{r^{N-2},r^{N-1}} G_{s^{N-1}}\cdot 1 \n
&\!\!\!\!\!\!\!\!&
= C^-_\lambda \oint\prod_{a=1}^{N-1} {\res{x^a}j{r^a}}\cdot
{\tDab 01}\prod_{a=2}^{N-1} {\Dab {a-1}a}^{-1/\beta}\cdot
\prod_{a=1}^{N-1} {\Daa a}^{1/\beta} 
\prod_{j=1}^{r^a}\left({\x aj}\right)^{s^a},
\label{e6.10}
\ea
\be
C^-_\lambda
={\prod_{a=2}^{N-1}
	\langle J_{{\laa a}},J_{{\laa a}}\rangle_{1/\beta} \over
  \prod_{a=1}^{N-1}r^a!
	\langle J_{{\laa a}},J_{{\laa a}}{\sRangle{1/\beta}{r^a}} }.
\ee

\no
If we replace respectively ${\x 0i}$ and ${\x ai}$ $(a\neq 0)$
with $x_i$ and $-1/z^a_i$,
then this formula also reduces to
the our integral formula \eq{J-vrvs}.

It is obvious that we may get various types of decomposition
of the Young diagram into the rectangles.
For example, we may obtain the (generalized) hook
decomposition if we use
only $\widetilde N^{(1/\beta)}_{r^0,r^1}$.
%
%
%
%
%

\section{Skew-Jack polynomials}


To calculate higher point dynamical correlation functions,
we will need the inner product 
$\langle J_{\lambda},J_{\mu}J_{\nu}\rangle^{\prime}$ or
equivalently the branching rule
\be
  J_{\mu}(x;\beta)J_{\nu}(x;\beta)
  =
  {\sum_{\lambda}}'C_{\mu\nu}^{\lambda}(\beta)J_{\lambda}(x;\beta),
\ee
where ${\sum_{\lambda}}'=\sum_{\lambda}
\langle J_{\lambda},J_{\lambda}\rangle^{-1}$.
This information is encoded in the skew-Jack polynomial
$J_{\lambda/\mu}(x;\beta)$ characterized 
by the following three equivalent definitions 
$$\begin{array}{lrl}
\mbox{(\romannumeral1)}\qquad &
  J_{\lambda/\mu}(x;\beta) 
  = &\!
  {\sum_{\nu}}'C_{\mu\nu}^{\lambda}(\beta)J_{\nu}(x;\beta),
  \label{sJ1}\\
\mbox{(\romannumeral2)}\qquad &
  \langle J_{\lambda},J_{\mu}J_{\nu}\rangle_{\beta}
  = &\!
  \langle J_{\lambda/\mu},J_{\nu}\rangle_{\beta}
  =C_{\mu\nu}^{\lambda}(\beta) \quad
  (\forall \nu),
  \label{sJ2}\\
\mbox{(\romannumeral3)}\qquad &
  {\sum_{\lambda}}'J_{\lambda/\mu}(x)J_{\lambda}(y)
  = &\!
  {\sum_{\nu}}'J_{\nu}(x)J_{\nu}(y)J_{\mu}(y).
  \label{sJ3}
\end{array}
$$


In order to match the inner products on the boson Fock space
and the ring of symmetric functions,
we rescale oscillators 
$a_n=\sqrt{2\beta}a'_n$ and $a_{-n}=\sqrt{\beta/2}a'_{-n}$, namely
\be
  [a'_n,a'_m]=\frac{n}{\beta}\delta_{n+m,0},
\ee
and define $\dagger$ operation by $a^{\prime\dagger}_n=a'_{-n}$.
We set 
$\phi'_-(z)=\sum_{n>0}\frac{1}{n}a'_{-n}z^n$ and
$\phi^{\prime\dagger}_-(z)=\sum_{n>0}\frac{1}{n}a'_nz^n$.
The correspondence between a state 
$\ket{f}=\hat{f}\ket{0}\in{\cal F}_0$ 
(or $\bra{f}=\bra{0}\hat{f}^{\dagger}$) and 
a symmetric function $f(x)$ is
\ba
  \ket{f}\mapsto f(x)
  &\!\!=\!\!&
  \bra{0}C'_{\beta}(x)\ket{f}, \quad
  C'_{\beta}(x)=e^{\beta\sum_{n>0}\frac{1}{n}a'_np_n}
  \bigl(=C_{\beta'}\bigr) \n
  &\!\!=\!\!&
  \bra{f}C^{\prime\dagger}_{\beta}(x)\ket{0}, \quad
  C^{\prime\dagger}_{\beta}(x)=e^{\beta\sum_{n>0}\frac{1}{n}a'_{-n}p_n}.
\ea
In these notations, the two inner products agrees,
\be
  \langle f,g\rangle_{\beta}=\langle f\ket{g}.
\ee
We remark that
\be
  \ket{fg}=\hat{f}\hat{g}\ket{0}
  \mapsto
  f(x)g(x)=\bra{0}C'_{\beta}(x)\ket{fg}.
  \label{fg}
\ee

For the Young diagram 
$\lambda'=((r^1)^{s^1},(r^2)^{s^2},\cdots,(r^{N-1})^{s^{N-1}})$,
we define
\ba
  \ket{J_{\lambda}}
  &\!\!=\!\!&
  \oint\prod_{j=1}^{r^1}\frac{dz^1_j}{2\pi iz^1_j}\cdot
  f^\pm_{\lambda}(z^1)
  \prod_{j=1}^{r^1}e^{\gamma_\pm\phi'_-(z^1_j)}\ket{0},
  \label{ket1} \\
  f^\pm_{\lambda}\left({1\over z^1}\right)
  &\!\!=\!\!& C^\pm_\lambda
  \oint\prod_{a=2}^{N-1} \prod_j^{r^a} \frac{dz^a_j}{2\pi i z^a_j}
                         \Gamma(z^{a-1},z^a)^{-\beta^{\pm 1}}\cdot
  \prod_{a=1}^{N-1}\Delta(z^a)^{\beta^{\pm 1}}
                   \prod_{i=1}^{r^a} (z^a_i)^{s^a},
\ea
with $\gamma_+ = \beta$ and $\gamma_- = -1$.

Then Jack and skew-Jack polynomials are given by
\ba
  J_{\lambda}(x;\beta)
  &\!\!=\!\!&
  \bra{0}C'_{\beta}(x)\ket{J_{\lambda}}
  =\bra{J_{\lambda}}C^{\prime\dagger}_{\beta}(x)\ket{0}, \\
  J_{\lambda/\mu}(x;\beta)
  &\!\!=\!\!&
  \bra{J_{\mu}}C'_{\beta}(x)\ket{J_{\lambda}}
  =\bra{J_{\lambda}}C^{\prime\dagger}_{\beta}(x)\ket{J_{\mu}}.
\ea
The former is proved in sections 5 and 6.
The latter is done as follows;
\ba
  \langle J_\lambda,J_\mu J_\nu\rangle_\beta
  &\!\!=\!\!&
  \langle J_\lambda\ket{J_\mu J_\nu}
  =\langle J_\lambda|\hat{J}_\mu\ket{J_\nu} \n
  &\!\!=\!\!&
  \langle\;\langle J_\lambda|\hat{J}_\mu C^{\prime\dagger}_{\beta}(x)
  \ket{0},
  \langle 0|C'_{\beta}(x)\ket{J_\nu}\;\rangle_\beta \n
  &\!\!=\!\!&
  \langle\;\langle J_\lambda|C^{\prime\dagger}_{\beta}(x)\ket{J_\mu},
  \langle 0|C'_{\beta}(x)\ket{J_\nu}\;\rangle_\beta \n
  &\!\!=\!\!&
  \langle J_{\lambda/\mu},J_\nu\rangle_\beta.
\ea
Here we have used \eq{fg} and 
$[\hat{J}_\mu,C^{\prime\dagger}_{\beta}(x)]=0$. $\Box$ \\
We can give another proof;
\ba
  {\sum_{\lambda}}'J_{\lambda/\mu}(x)J_{\lambda}(y)
  &\!\!=\!\!&
  {\sum_{\lambda}}'
  \bra{0}C'_{\beta}(y)\ket{J_{\lambda}}
  \bra{J_{\lambda}}C^{\prime\dagger}_{\beta}(x)\ket{J_{\mu}} 
  =
  \bra{0}C'_{\beta}(y)C^{\prime\dagger}_{\beta}(x)\ket{J_{\mu}} \n
  &\!\!=\!\!&
  \bra{0}C'_{\beta}(y)\ket{J_{\mu}}
  \prod_i\prod_j(1-x_iy_j)^{-\beta} \n
  &\!\!=\!\!&
  {\sum_{\nu}}'J_{\nu}(x)J_{\nu}(y)J_{\mu}(y).
\ea
Here we have used \eq{e6.3} and
completeness of $\{\ket{J_{\lambda}}\}$ in ${\cal F}_0$. $\Box$

Now we can write down integral representations of
skew-Jack polynomials.
By using \eq{ket1}, we have
\def\eMu{a} \def\eLa{b}
\be
  J_{\lambda/\mu}(x;\beta)
  =
  \oint\prod_{j=1}^{\ell(\mu)}\frac{dw^1_j}{2\pi iw^1_j}
       \prod_{j=1}^{\ell(\lambda)}\frac{dz^1_j}{2\pi iz^1_j}\cdot
  f^{\eMu}_{\mu}(w^1)  f^\eLa_{\lambda}(z^1)
  \Gamma(w^1,z^1)^{-\gamma_{\eMu}\gamma_\eLa\over\beta}
  \Gamma(x,z^1)^{-\gamma_\eLa}.
\label{SkewJack}
\ee

More generally, the skew-Jack polynomial can be written 
in the integral transformation $N_{N,M}^{(\beta)}$ in \eq{IntTrans} or 
in the power-sum as follows:
\ba
  J_{\lambda/\mu}(x_1,\cdots,x_N)
  &\!\!=\!\!&
  {\langle J_\lambda,J_\lambda\rangle_\beta \over 
 M!\langle J_\lambda,J_\lambda\rangle'_{\beta;M} }
 N_{N,M}^{(\beta)} 	J_\lambda(t_1,\cdots,t_M) 
		J_\mu\left({1\over t_1},\cdots,{1\over t_M}\right) \n
  J_{\lambda/\mu}(p)
  &\!\!=\!\!&
 {J}_\mu(\overline{p}) J_\lambda(p) \cdot 1,
\ea
for all $M\geq\ell(\lambda)$.
Here $\overline p_n = {n\over\beta}{\partial\over\partial p_n}$.
When $M=\ell(\lambda)$, this reduces to \eq{SkewJack}.

\section{Discussion}

There are some points which may be interesting
if it is clarified in the future study.

\begin{enumerate}


\item
The methods in section 6 and 7 are applicable to the $q$--analogue
of the Jack polynomial, that is the Macdonald polynomial.
Integral representation of the (skew-)Macdonald polynomial can be
obtained from that of the Jack polynomial by replacing $\Delta$, 
$\Gamma$ and $\widetilde{\Gamma}$ with $q$--deformed ones \cite{rAOS}.


\item
It is challenging to calculate dynamical
$n$--point correlation functions  for $n$ greater than two.
For this purpose, we will have to use the inner-product
$\langle J_{\lambda},J_{\mu}J_{\nu}\rangle^{\prime}$,
whose integral representation has been obtained in section 7.


\item
In our construction, the choice of the
$W$ algebra depends on the form of the Young diagram.
It will be natural to conjecture that there is a
underlying symmetry which explains the appearance
of various symmetries.


\item
This is also related with Kac-Moody algebras.
In fact, the operator $C_{\beta '}$ in \eq{sfcN}
corresponds to the product of $N$--vertex operators of $\widehat{sl(N)}$
with fundamental representations.
The level is $k+N = 1/\beta$.
Hence if we decompose the Young diagram as $r^a = N-a$
and allow $s^a$'s vanish,
then the integrand of \eq{J+vrvs} is just
the $\phi$--boson part of that of a zero-weight $N$--point function.


\item
The large distance behavior of correlation functions
is described by $c=1$ CFT \cite{rKY}.
To give the eigenfunction, as we observed, $c<1$ CFT plays the essential
role.  It is interesting to understand the relation
between them.


\item The situation that the only relevant
states are given by the null states reminds us of
the situation in the quantum gravity \cite{rLZ}.
This fact comes from the similarity between our Hamiltonian
\eq{eHb} and the BRST currents when we replaced
the $a_{-n}$ by the ghost field.


\item
Another analogy with the gravity is that our Hamiltonian
is a deformation of that of quantum gravity
considered by Ishibashi and Kawai \cite{rIK}.
It may be interesting to understand what
quantum gravity system our Hamiltonian (or its continuum
limit) describes.
It is well known that the matrix model
can be described by free fermions.  Our construction
shows that we may define similar model even if we
replace these fermions with anyons.


\item
One can generalize our state--function correspondence \eq{sfcN}
to a invertible map.
In fact, if we introduce $N-1$ kinds of power-sums
$p_n^{(a)}$ $(a=1,\cdots,N-1)$,
then the operator
$C_{\beta'}\equiv 
\exp\left\{ \beta' \sum_{n>0}{1\over n} \sum_{b=1}^{N-1}
\vec\Lambda_b\cdot\vec a_n \, p_n^{(b)} \right\}$
gives such a map.
When $s^a = \beta(1-r^a+r^{a-1})-1$ with $r^0 = 0$,
the function
$Z(p)\equiv
\bra{\vec{\lambda}_{\vec r,\vec s}^+}C_{\beta'}\ket{\chi_{\vec r,\vec s}^+}$
is regarded as a generalized partition function
of the conformal matrix model in \cite{rKMMM} of $\beta=1$.
However, it is still unknown what system this map describes in general.


\item The two--point function derived by Ha \cite{rHa}
has following form,
\be
\langle 0 | \rho(x,t) \rho(0,0) |0\rangle
=
C \prod_{i=1}^q\int_0^\infty dx_i
\prod_{j=1}^p \int_0^1 dy_i Q^2
F(q,p,\beta|{\{ x_i, y_j\}}) \cos(Qx)e^{-iEt},
\ee
where $Q=2\pi\rho_0(\sum_{i=1}^q x_i+\sum_{j=1}^p y_j)$,
 $E=(2\pi\rho_0)^2(\sum_{i=1}^q \epsilon_P(x_i)+
\sum_{j=1}^p \epsilon_H(y_j))$. $\epsilon_P$ and
$\epsilon_H$ are the energy for pseudo-particle
and hole.
The form factor is given by,
\ba
F(q,p,\beta|\{ x_i, y_j\})&=&
\prod_{i=1}^q \prod_{j=1}^p (x_i+\beta y_j)^{-2}
\frac{\prod_{i<j} (x_i-x_j)^{2\beta}
\prod_{i<j}(y_i-y_j)^{2/\beta}}{
\prod_{i=1}^q \epsilon_P({x_i})^{1-\beta}
\prod_{j=1}^p \epsilon_H({y_j})^{1-1/\beta}}\n
& \propto &
\frac{ \langle
\prod_{i=1}^q :e^{\alpha_+ \phi(x_i/\alpha_+)}:
\prod_{j=1}^p :e^{\alpha_- \phi(y_j/\alpha_-)}:\rangle}{
\prod_{i=1}^q \epsilon_P({x_i})^{1-\beta}
\prod_{j=1}^p \epsilon_H({y_j})^{1-1/\beta}}.
\ea
The Vertex operators in the final expression
are nothing but the screening charges we used
in section 4 and 5 \cite{rKhv}.  Although there is a definite gap
between our approach and theirs,
this fact may indicate that some structure we obtained
in this paper survives in the thermo-dynamical limit.


\end{enumerate}

\vskip 5mm
\noindent{\bf Acknowledgments:}

We would like to thank
M.~R.~Douglas,
V.~Kac, T.~Kawai, A.~Kirrilov Jr.,
A.~Matsuo, E.K.~Sklyanin, Y.~Yamada and S.~K.~Yang
for discussions. 		
Y.M. and S.O. would like to thank members of Kyung Hee University
for their hospitality.
This work is supported in part by Grant-in-Aid for Scientific
Research from Ministry of Science and Culture.


\section*{Appendix A : Construction of states with a excitation of
$M$ pseudo-particle (pseudo-hole)}
\setcounter{section}{1}
\renewcommand{\thesection}{\Alph{section}}

\subsection{}
Action with $\hat{H}_{\beta}$ on
$\prod_{i=1}^Me^{\alpha_{\pm}\phi_-(z_i)}\ket{\alpha}$
($|z_1|>\cdots>|z_M|$) is given by \cite{rAMOS}
\be
  \hat{H}_{\beta}\prod_{i=1}^Me^{\alpha_{\pm}\phi_-(z_i)}\ket{\alpha}
  =
  H'_{\beta,M}\prod_{i=1}^Me^{\alpha_{\pm}\phi_-(z_i)}\ket{\alpha},
  \label{HM}
\ee
where
\ba
  H'_{\beta,M}
  &\!\!=\!\!&
  \sum_{i=1}^M\biggl(
  \frac{\sqrt{\beta}}{\alpha_{\pm}/\sqrt{2}}D_i^2
  +\Bigl(N_0\beta-\sqrt{\beta}\sfrac{\alpha_{\pm}}{\sqrt{2}}\Bigr)D_i
  \biggr)
  +2\sqrt{\beta}\sfrac{\alpha_{\pm}}{\sqrt{2}}
  \sum_{i,j=1 \atop i<j}^M\frac{1}{1-\frac{z_j}{z_i}}
  \Bigl(\sfrac{z_j}{z_i}D_i-D_j\Bigr) \n
  &\!\!=\!\!&
  \frac{\sqrt{\beta}}{\alpha_{\pm}/\sqrt{2}}
  H_{\frac{1}{2}\alpha_{\pm}^2}
  +\Bigl(N_0\beta-M\sqrt\beta\sfrac{\alpha_{\pm}}{\sqrt{2}}
  \Bigr)P.
\ea
Here $D_i=z_i\frac{\partial}{\partial z_i}$,
and $H$ and $P$ are the Hamiltonian
and momentum with variable $z_i$ ($i=1,\cdots,M$).
Expanding this, we obtain
\ba
  &&
  \hat{H}_{\beta}
  \hat{J}^{\pm}_{n_1}\cdots\hat{J}^{\pm}_{n_M}\ket{\alpha} \n
  &\!\!=\!\!&
  \sum_{i=1}^M\biggl(
  \frac{\sqrt{\beta}}{\alpha_{\pm}/\sqrt{2}}n_i^2
  +\Bigl(N_0\beta-\sqrt{\beta}\sfrac{\alpha_{\pm}}{\sqrt{2}}
  (2i-1)\Bigr)n_i\biggr)
  \hat{J}^{\pm}_{n_1}\cdots\hat{J}^{\pm}_{n_M}\ket{\alpha} \n
  &&
  +2\sqrt{\beta}\sfrac{\alpha_{\pm}}{\sqrt{2}}
  \sum_{i,j=1 \atop i<j}^M\sum_{\ell=1}^{n_j}(n_i-n_j+2\ell)
  \hat{J}^{\pm}_{n_1}\cdots\hat{J}^{\pm}_{n_i+\ell}\cdots
  \hat{J}^{\pm}_{n_j-\ell}\cdots\hat{J}^{\pm}_{n_M}\ket{\alpha}.
\ea
The subspace with $\hat{P}=|\lambda|$ has
basis $\hat{J}^{\pm}_{n_1}\cdots\hat{J}^{\pm}_{n_{|\lambda|}}
\ket{\alpha}$ ($n_1\geq\cdots\geq n_{|\lambda|}\geq 0$,
$\sum_in_i=|\lambda|$),
on which $\hat{H}_{\beta}$ is represented as a triangular matrix.
The energy eigenvalue $\epsilon_{\beta,\lambda}$ \eq{ebl} can be
read from the diagonal elements.
By diagonalizing this triangular matrix,
the eigenstates are determined as
\ba
  \ket{J_{\lambda}^{\pm}}
  &\!\!=\!\!&
  \biggl(\prod_{i=1}^M \hat{J}^{\pm}_{\lambda_i}
  +\cdots \,\biggr)\ket{\alpha}, \quad
  (\lambda=(\lambda_1,\cdots,\lambda_M)), \n
  \hat{H}_{\beta}\ket{J_{\lambda}^{\pm}}
  &\!\!=\!\!&
  \epsilon_{\beta,\lambda^{\pm}}\ket{J_{\lambda}^{\pm}},\qquad
  (\lambda^+=\lambda,\lambda^-=\lambda').
\ea
For example, $M=2$ case was explicitly solved,
\ba
  \ket{J_{(\lambda_1,\lambda_2)}^{\pm}}
  &\!\!=\!\!&
  \sum_{\ell=0}^{\lambda_2}c^{\pm}(\lambda_1-\lambda_2,\ell)
  \hat{J}_{\lambda_1+\ell}^{\pm}\hat{J}_{\lambda_2-\ell}^{\pm}
  \ket{\alpha}, \n
  c^{\pm}(\lambda,\ell)
  &\!\!=\!\!&
  \frac{\lambda+2\ell}{\lambda+\ell}
  \prod_{j=1}^{\ell}\frac{\lambda+j}{j}\cdot
  \prod_{i=1}^{\ell}
  \frac{-\sqrt{\beta}\sfrac{\alpha_{\pm}}{\sqrt{2}}
	+\frac{\sqrt{\beta}}{\alpha_{\pm}/\sqrt{2}}(i-1)}
       { \sqrt{\beta}\sfrac{\alpha_{\pm}}{\sqrt{2}}
	+\frac{\sqrt{\beta}}{\alpha_{\pm}/\sqrt{2}}(\lambda+i)}.
  \label{cM2}
\ea
The Jack polynomial is obtained by the state--function correspondence
\eq{sfc},
$J_{\lambda}^{\pm}(x)=\bra{\alpha}C_{\beta'}\ket{J_{\lambda}^{\pm}}$.
The normalization between $J_{\lambda}^{\pm}(x)$ and $J_{\lambda}(x)$
are given by \cite{rS}
\ba
  J_{\lambda}^+(x;\beta)
  ={\cal N}_{\lambda}^+ J_{\lambda}(x;\beta), &&
  {\cal N}_{\lambda}^+
  =
  \prod_{s\in\lambda}
  \frac{(\ell_{\lambda}(s)+1)\beta+a_{\lambda}(s)}
       {\ell_{\lambda}(s)\beta+a_{\lambda}(s)+1},
\label{eJplus}\\
  J_{\lambda}^-(x;\beta)
  ={\cal N}_{\lambda}^- J_{\lambda'}(x;\beta), &&
  {\cal N}_{\lambda}^- = (-1)^{|\lambda|}.
\label{eJminus}\ea

\subsection{}

Instead of using mode expansion,
next we will consider the eigenstates in the following form,
\be
  \ket{J_{\lambda}^{\pm}}
  =
  \oint\prod_{j=1}^M\frac{dz_j}{2\pi iz_j}\cdot
  \prod_{i=1}^Mz_i^{-\lambda_i}\cdot f_{\lambda}^{\pm}(z_1,\cdots,z_M)
  \prod_{i=1}^Me^{\alpha_{\pm}\phi_-(z_i)}\ket{\alpha}.
\ee
Here the integration contour is shown in Figure 2(a).
{}From the argument of mode expansion,
$f_{\lambda}^{\pm}(z_1,\cdots,z_M)$ is a finite sum of
terms $\prod_{i<j}(\frac{z_j}{z_i})^{n_{ij}}$ ($n_{ij}\geq 0$) and
has a constant term $1$.
In the case of $\beta=1$ when the Jack polynomial reduces to
the Schur polynomial, $f_{\lambda}^{\pm}$ is independent
of $\lambda$ and given by
$f_{\lambda}^{\pm}(z_1,\cdots,z_M)=\prod_{i<j}(1-\frac{z_j}{z_i})$,
which can be written in a determinant form,
$\hat{J}_{\lambda}^{\pm}=\det
(\hat{J}_{\lambda_i-i+j}^{\pm})_{1\leq i,j\leq M}$.

Although $f_{\lambda}^{\pm}$ has only finite number of terms,
we can add to $f_{\lambda}^{\pm}$ the terms that do not
contribute to the integral.
This freedom may give us the possibility to write down solutions
for general $M$.
In fact integral representations of such solutions are given in section 5,6.
In the following, we allow that $f_{\lambda}^{\pm}$ may have
infinitely many terms.
We will give a sufficient condition for such $f_{\lambda}^{\pm}$.
By setting
$f_{\lambda}^{\pm}\prod_{i}z_i^{-\lambda_i}
=F_{\lambda}^{\pm}\prod_{i<j}(z_i-z_j)^{\alpha_{\pm}^2}$,
$\ket{J_{\lambda}^{\pm}}$ is rewritten as
\ba
  \ket{J_{\lambda}^{\pm}}
  &\!\!=\!\!&
  \oint\prod_{j=1}^M\frac{dz_j}{2\pi iz_j}\cdot
  \prod_{i,j=1 \atop i<j}^M(z_i-z_j)^{\alpha_{\pm}^2}\cdot
  F_{\lambda}^{\pm}(z_1,\cdots,z_M)
  \prod_{i=1}^Me^{\alpha_{\pm}\phi_-(z_i)}\ket{\alpha} \n
  &\!\!=\!\!&
  \oint\prod_{j=1}^M\frac{dz_j}{2\pi iz_j}\cdot
  F_{\lambda}^{\pm}(z_1,\cdots,z_M)
  \prod_{i=1}^Mz_i^{-\alpha_{\pm}(\alpha-M\alpha_{\pm})}
  \prod_{i=1}^M:e^{\alpha_{\pm}\phi(z_i)}:\ket{\alpha-M\alpha_{\pm}},
  \label{JF}
\ea
where we may have to choose appropriate integration contour.
By using \eq{HM} and integration by parts,
\begin{eqnarray*}
  &&
  \hat{H}_{\beta}\ket{J_{\lambda}^{\pm}}=
  \epsilon_{\beta,\lambda^{\pm}}\ket{J_{\lambda}^{\pm}} \\
  &\!\!=\!\!&
  \oint\prod_{j=1}^M\frac{dz_j}{2\pi iz_j}\cdot
  \prod_{i,j=1 \atop i<j}^M(z_i-z_j)^{\alpha_{\pm}^2}\cdot
  F_{\lambda}^{\pm}(z_1,\cdots,z_M)
  H'_{\beta,M}
  \prod_{i=1}^Me^{\alpha_{\pm}\phi_-(z_i)}\ket{\alpha} \\
  &\!\!=\!\!&
  \oint\prod_{j=1}^M\frac{dz_j}{2\pi iz_j}\cdot
  (H'_{\beta,M})^{\dagger}\biggl(
  \prod_{i,j=1 \atop i<j}^M(z_i-z_j)^{\alpha_{\pm}^2}\cdot
  F_{\lambda}^{\pm}(z_1,\cdots,z_M)
  \biggr)\cdot
  \prod_{i=1}^Me^{\alpha_{\pm}\phi_-(z_i)}\ket{\alpha},
\end{eqnarray*}
we obtain the sufficient condition for $F_{\lambda}^{\pm}$ ;
this $F_{\lambda}^{\pm}(z_1,\cdots,z_M)$ is homogeneous and
the eigenfunction of $H$ with variable $z_i$,
\be
  H_{\frac{1}{2}\alpha_{\pm}^2}F_{\lambda}^{\pm}
  =
  \epsilon_{\frac{1}{2}\alpha_{\pm}^2,\tilde{\lambda}}
  F_{\lambda}^{\pm},\quad
  \tilde{\lambda}_i=-\lambda_i-(M-i)\alpha_{\pm}^2.
\ee
Note a symmetry
$\epsilon_{\beta,\lambda} = \epsilon_{\beta,\overline\lambda}$ with
$\overline\lambda_i = -\lambda_i + \beta(2i-1)$.
Also note that $F_{\lambda}^{\pm}$ may not be a polynomial.

We remark that the following property of $H_{\beta}$.
Let $\psi_{\lambda}(x_1,\cdots,x_{N_0})$, which is symmetric
and homogeneous, be the eigenfunction of $H_{\beta}$,
$H_{\beta}\psi_{\lambda}=\epsilon_{\beta,\lambda}\psi_{\lambda}$.
Since the original Hamiltonian $H_{CS}$
is symmetric with respect to the transformation
$\beta\leftrightarrow 1-\beta$,
the function $\tilde{\psi}_{\tilde{\lambda}}=\psi_{\lambda}
\prod_{i<j}(x_i-x_j)^{2\beta-1}$ is also the eigenfunction of $H$,
\be
  H_{\tilde{\beta}}\tilde{\psi}_{\tilde{\lambda}}
  =
  \epsilon_{\tilde{\beta},\tilde{\lambda}}
  \tilde{\psi}_{\tilde{\lambda}},\quad
  \tilde{\beta}=1-\beta,\quad
  \tilde{\lambda}_i=\lambda_i+(2\beta-1)(N_0-i).
\ee

Setting $\tilde{F}_{\lambda}^{\pm}=
F_{\lambda}^{\pm}\prod_{i<j}(z_i-z_j)^{\alpha_{\pm}^2-1}$,
then $\ket{J_{\lambda}^{\pm}}$ is rewritten as
\be
  \ket{J_{\lambda}^{\pm}}
  =
  \oint\prod_{j=1}^M\frac{dz_j}{2\pi iz_j}\cdot
  \prod_{i,j=1 \atop i<j}^M(z_i-z_j)\cdot
  \tilde{F}_{\lambda}^{\pm}(z_1,\cdots,z_M)
  \prod_{i=1}^Me^{\alpha_{\pm}\phi_-(z_i)}\ket{\alpha}.
\ee
Due to above property, the sufficient condition is reexpressed
as follows;
$\tilde{F}_{\lambda}^{\pm}(z_1,\cdots,z_M)$ is homogeneous and
the eigenfunction of $H$ with variable $z_i$,
\be
  H_{1-\frac{1}{2}\alpha_{\pm}^2}\tilde{F}_{\lambda}^{\pm}
  =
  \epsilon_{1-\frac{1}{2}\alpha_{\pm}^2,\tilde{\lambda}}
  \tilde{F}_{\lambda}^{\pm},\quad
  \tilde{\lambda}_i=-\lambda_i-(M-i).
  \label{tF}
\ee

For $M=2$, we can easily find a solution of this equation.
Let us set $\tilde{F}_{\lambda}^{\pm}=\tilde{f}_{\lambda}^{\pm}
\prod_{i}z_i^{-\lambda_i-(M-i)}$.
Substituting the form
$\tilde{f}_{(\lambda_1,\lambda_2)}^{\pm}(z_1,z_2)
=\sum_{n=0}^{\infty}c_n^{\pm}(\frac{z_2}{z_1})^n$ ($c_0^{\pm}=1$)
in \eq{tF},
we obtain the result that $\tilde{f}_{(\lambda_1,\lambda_2)}^{\pm}$ is
given by the hypergeometric function,
\be
  \tilde{f}_{(\lambda_1,\lambda_2)}^{\pm}(z_1,z_2)
  =
  {}_2F_{\,1} \biggl[
  \begin{array}{c}
    \lambda_1-\lambda_2+1,\;-\frac{1}{2}\alpha_{\pm}^2+1 \\
    \lambda_1-\lambda_2+\frac{1}{2}\alpha_{\pm}^2+1
  \end{array}
  ;\frac{z_2}{z_1} \biggr].
\ee
$f_{(\lambda_1,\lambda_2)}^{\pm}(z_1,z_2)=
(1-\frac{z_2}{z_1})\tilde{f}_{(\lambda_1,\lambda_2)}^{\pm}(z_1,z_2)$
agrees with the generating function for $c^{\pm}(\lambda,\ell)$ of
\eq{cM2},
\be
  \sum_{l=0}^{\infty}
  c^{\pm}(\lambda_1-\lambda_2,\ell)\Bigl(\frac{z_2}{z_1}\Bigr)^{\ell}
  =
  f_{(\lambda_1,\lambda_2)}^{\pm}(z_1,z_2).
\ee

For $M=3$, we can derive the recurrence formula for the
coefficients of the Taylor series of $\tilde{F}_{\lambda}^{\pm}$,
which has a property of the root system of $sl_3$.
However it is hard to obtain a general solution for
$\tilde{F}_{\lambda}^{\pm}$.



\section*{Appendix B}

We consider the analytic continuation of the following
integral \cite{rDFF,rTK},
\be
  I=
  \int\prod_{j=1}^rdz_j\cdot
  \prod_{i,j=1 \atop i<j}^r(z_i-z_j)^{2\alpha}\cdot
  \prod_{i=1}^rz_i^{\alpha'}\cdot
  g(z_1,\cdots,z_r).
\ee
Let $I_1$ to be $I$ with the integration region $0<z_r<\cdots<z_1<1$,
and $I_2$ to be $I$ with the integration contour shown in Figure 2(b).
We assume that $g(z_1,\cdots,z_r)$ has no poles at $z_i=z_j$.
Then $I_1$ and $I_2$ are related as follows:
\be
  I_2
  =
  (-1)^r\prod_{j=1}^r(1-a'a^{j-1})\cdot
  \prod_{j=1}^r(1+a+\cdots+a^{j-1})\cdot
  I_1,
\ee
where $a$ and $a'$ are
\be
  a=e^{2\pi i\alpha},\quad a'=e^{2\pi i\alpha'}.
\ee
Therefore we have
\be
  \frac{1}{(2\pi i)^r}I_2
  =
  e^{\pi ir((r-1)\alpha+\alpha')}
  \prod_{j=1}^r\frac{\sin\pi j\alpha}{\sin\pi\alpha}\,
  \frac{\sin\pi((j-1)\alpha+\alpha')}{\pi}
  \cdot I_1.
  \label{I2I1}
\ee

When
$g(z_1,\cdots,z_r)=\prod_{i=1}^r(1-z_i)^{\alpha^{\prime\prime}}$,
this integral is known as the Selberg integral \cite{rSel},
\be
  I_1
  =
  \prod_{j=1}^r
  \frac{\Gamma(j\alpha)}{\Gamma(\alpha)}\,
  \frac{\Gamma((j-1)\alpha+\alpha'+1)
	\Gamma((j-1)\alpha+\alpha^{\prime\prime}+1)}
       {\Gamma((r-2+j)\alpha+\alpha'+\alpha^{\prime\prime}+2)}.
  \label{SI}
\ee

The normalization constant ${\cal N}_{r,s}^+$ in \eq{J+rs} is
given by the coefficient of $(J_s^+)^r$ in
$\bra{\alpha_{r,s}}C_{\beta'}\ket{\chi_{r,s}^+}$,
\be
  {\cal N}_{r,s}^+
  =
  \oint\prod_{j=1}^r\frac{dz_j}{2\pi iz_j}\cdot
  \prod_{i,j=1 \atop i<j}^r(z_i-z_j)^{2\beta}\cdot
  \prod_{i=1}^rz_i^{(1-r)\beta},
\ee
where the integration contour is shown in Figure 2(b).
Using \eq{I2I1} and \eq{SI}, setting $\alpha=\beta$,
$\alpha^{\prime\prime}=0$ and tending to a limit 	
$\alpha'\rightarrow(1-r)\beta-1$, we obtain
\be
  {\cal N}_{r,s}^+
  =
  \frac{1}{r!}
  \prod_{j=1}^r\frac{\sin\pi j\beta}{\sin\pi\beta}\cdot
  \frac{\Gamma(r\beta+1)}{\Gamma(\beta+1)^r}.
\ee
Here we have used $\Gamma(z)\Gamma(1-z)=\frac{\pi}{\sin\pi z}$.
The other one ${\cal N}_{r,s}^-$ is obtained from ${\cal N}_{r,s}^+$ by
replacing $\beta$, $r$, $s$ with $\frac{1}{\beta}$, $s$, $r$.

The normalization constant ${\cal N}_{\vec{r},\vec{s}}^+$ in
\eq{J+vrvs} is given by the coefficient of
$(J_{s^1+\cdots+s^{N-1}}^+)^{r^{N-1}}$
$(J_{s^1+\cdots+s^{N-2}}^+)^{r^{N-2}-r^{N-1}}\cdots$
$(J_{s^1}^+)^{r^1-r^2}$ in
$\bra{\vec{\lambda}_{\vec{r},\vec{s}}^+}C_{\beta'}
 \ket{\chi_{\vec{r},\vec{s}}^+}$,
\ba
  {\cal N}_{\vec{r},\vec{s}}^+
  &\!\!=\!\!&
  \oint\prod_{a=1}^{N-1}\prod_{j=1}^{r^a}\frac{dz^a_j}{2\pi iz^a_j}\cdot
  \prod_{a=1}^{N-1}\prod_{i,j=1 \atop i<j}^{r^a}
  (z^a_i-z^a_j)^{2\beta}\cdot
  \prod_{a=1}^{N-2}\prod_{i=1}^{r^a}\prod_{j=1}^{r^{a+1}}
  (z^a_i-z^{a+1}_j)^{-\beta} \n
  &&
  \times\prod_{a=1}^{N-1}\prod_{j=1}^{r^a}
  (z^a_j)^{(1-r^a+r^{a+1})\beta-s^a}\cdot
  \prod_{a=1}^{N-1}\frac{1}{(r^a-r^{a+1})!}\cdot
  \sum_{\sigma}\prod_{j=1}^{r^1}(z^1_j)^{\lambda_{\sigma(j)}},
\ea
where $\sum_{\sigma}$ stands for the summation over all permutations
of $r^1$ objects and $\lambda_i$ is
$(\lambda_1,\cdots,\lambda_{r^1})=
((s^1+\cdots+s^{N-1})^{r^{N-1}},
(s^1+\cdots+s^{N-2})^{r^{N-2}-r^{N-1}},\cdots,(s^1)^{r^1-r^2})$.
The other one ${\cal N}_{\vec{r},\vec{s}}^-$ is obtained from
${\cal N}_{\vec{r},\vec{s}}^+$ by
replacing $\beta$, $\vec{r}$ and $\vec{s}$ with
$\frac{1}{\beta}$, $\vec{s}$ and $\vec{r}$, respectively.
At present we have not obtained the explicit form of
${\cal N}_{\vec{r},\vec{s}}^{\pm}$.
For a positive integer $\beta$, the normalization constants
are explicitly given in section 6, \eq{C+} and \eq{e6.11}.

\section*{Appendix C : Explicit examples}

The null state at level $n$ is defined as,
\be\label{eq:NSC}
L_n |\chi\rangle = 0\quad (n>0),\qquad
L_0 |\chi\rangle = (h+n)|\chi\rangle
\ee
By a standard argument, it has null norm with any states
in the Verma module,
\be\label{eq:NSC2}
\langle * | \chi \rangle =0.
\ee
The existence of such states depend crucially on the
choice of parameter $c$ and $h$.
Cerebrated Kac formula shows that if they are
explicitly parametrized as,
\be\label{eq:Kac}
c=1-\frac{6(\beta-1)^2}{\beta}\qquad
h_{rs}=\frac{(\beta r-s)^2-(\beta-1)^2}{4\beta},
\ee
for an arbitrary parameter $\beta$, there exists null state
at level $rs$.
Some of the lower lying states can be explicitly
obtained by solving the conditions \eq{eq:NSC}.
Let us introduce the notation $|\chi_{rs}\rangle$
as the null state that occurs in the highest module
over the vacuum $|h_{rs}\rangle$ at level $rs$.
We obtain,
\ba\label{eq:Lower}
|\chi_{11}\rangle&=& L_{-1}|h_{11}\rangle\n
|\chi_{21}\rangle&=& (L_{-2}-\frac{1}{\beta}L_{-1}^2)|h_{21}\rangle\n
|\chi_{12}\rangle&=& (L_{-2}-\beta L_{-1}^2)|h_{12}\rangle\n
|\chi_{31}\rangle&=& \left((1-2\beta) L_{-3}+
	2L_{-2}L_{-1}-\frac{1}{2\beta}L_{-1}^3\right)|h_{31}\rangle\n
|\chi_{13}\rangle&=& \left((1-2/\beta) L_{-3}+
	2L_{-2}L_{-1}-\frac{\beta}{2}L_{-1}^3\right)|h_{13}\rangle\n
|\chi_{41}\rangle&=& \left((1-4\beta+6\beta^2) L_{-4}
+\frac{5-12\beta}{3}L_{-3}L_{-1}
\right.\n
&&\left.
-\frac{3\beta}{2}L_{-2}^2
+\frac{5}{3}L_{-2}L_{-1}^2-\frac{1}{6\beta}L_{-1}^4\right)
|h_{41}\rangle\n
|\chi_{14}\rangle &=& \left. |\chi_{41}\rangle \right|_{\beta
\rightarrow 1/\beta,\ h_{41}\rightarrow h_{14}}\n
|\chi_{22}\rangle &=& \left(L_{-4}+
\frac{2(\beta^2-3\beta+1)}{3(\beta-1)^2}L_{-3}L_{-1}-
\frac{(\beta+1)^2}{3\beta}L_{-2}^2\right.\n
&&\left.+
\frac{2(\beta^2+1)}{3(\beta-1)^2}L_{-2}L_{-1}^2-
\frac{\beta}{3(\beta-1)^2}L_{-1}^4\right)
|h_{22}\rangle.
%
\ea
The duality $\beta \leftrightarrow 1/\beta$
is realized as a symmetry $r\leftrightarrow s$.

These null states can be regarded as the
Jack polynomial once we use the bosonic representation.
In mode expansion, Virasoro charges are replaced as,
\be\label{eq:coulomb}
L_n=\frac{1}{2}\sum_{m\in {\bf Z}} :a_{n+m} a_{-m}: -\alpha_0
 (n+1) a_n.
\ee
The central charge \eq{eq:Kac} is obtained by choosing
$\alpha_0 =\frac{1}{\sqrt{2}}({\sqrt{\beta}-\sqrt{1/\beta}})$.
The highest weight state $|h_{rs}\rangle$ can be replaced
by the Fock space vacuum $\ket{\alpha_{rs}}$,
with
$$
\alpha_{rs}=\frac{1}{\sqrt 2}\left((1+r)\sqrt \beta
-(1+s)\sqrt{1/\beta}\right).
$$
In terms of free boson oscillators,
the null states \eq{eq:Lower} are written as
(up to overall normalization),
\ba
|\chi_{11}\rangle & \sim & a_{-1} |\alpha_{11}\rangle \n
|\chi_{21}\rangle & \sim & \left(a_{-2}
	- \sqrt{2/\beta}a_{-1}^2\right)|\alpha_{21}\rangle \n
|\chi_{12}\rangle & \sim & \left(a_{-2}
	+ \sqrt{2\beta}a_{-1}^2\right)|\alpha_{12}\rangle \n
|\chi_{31}\rangle & \sim & \left(a_{-3}
	-\frac{3}{\sqrt{2\beta}}a_{-2}a_{-1}
	+\frac{1}{\beta} a_{-1}^3
	\right)|\alpha_{31}\rangle \n
|\chi_{13}\rangle & \sim & \left(a_{-3}
	+3{\sqrt{\frac{\beta}{2}}}a_{-2}a_{-1}
	+{\beta} a_{-1}^3
	\right)|\alpha_{13}\rangle \n
|\chi_{41}\rangle & \sim & \left(a_{-4}
	-\frac{8}{3\sqrt{2\beta}} a_{-3}a_{-1}
	-\frac{1}{\sqrt{2\beta}} a_{-2}^2
	+\frac{2}{\beta}a_{-2}a_{-1}^2
	-\frac{\sqrt{2}}{3\beta\sqrt{\beta}} a_{-1}^4
	\right)|\alpha_{41}\rangle \n
|\chi_{22}\rangle & \sim & \left( a_{-4}
	+\frac{4\sqrt{2\beta}}{1-\beta}
	 a_{-3}a_{-1}
	 -2\frac{1+\beta+\beta^2}{\sqrt{2\beta}(1-\beta)}a_{-2}^2
    -4 a_{-2}a_{-1}^2
	-\frac{2\sqrt{2\beta}}{1-\beta} a_{-1}^4
	\right)|\alpha_{22}\rangle.
\ea
The state $\ket{\chi_{s,r}}$ is obtained from $\ket{\chi_{r,s}}$
by $\beta\leftrightarrow 1/\beta$ and $a_n \leftrightarrow -a_n$.

To translate these expressions into symmetric functions,
one can apply the rule,
\be
a_{-n}\rightarrow \sqrt{\frac{\beta}{2}}p_n(x),
\qquad
\ket{\alpha_{rs}}\rightarrow 1,
\qquad
p_n=\sum_{i}(x_i)^n,
\ee
which gives the Jack polynomials for the rectangular Young diagram
$\lambda=\{ r^s\}$.

\vspace{10mm}



\end{document}